\documentclass[12pt]{article}

\usepackage[margin=1in]{geometry}
\usepackage{amsmath,amssymb,amsfonts}

\usepackage{natbib}
\usepackage{graphicx,xcolor,hyperref,url}
\usepackage{subcaption}

\newcommand{\Reals}{\mathbb{R}}
\newcommand{\Nats}{\mathbb{N}}
\newcommand{\Tr}{^{\rm{T}}}

\newcommand{\indsim}{\overset{ind}{\sim}}
\newcommand{\iidsim}{\overset{iid}{\sim}}

\newcommand{\response}{y}
\newcommand{\totalsize}{n}
\newcommand{\numgroups}{m}
\newcommand{\groupsize}{n}
\newcommand{\covdim}{p}
\newcommand{\numfun}{\covdim}
\newcommand{\fundim}{d}

\newcommand{\mean}{\mu}
\newcommand{\link}{g}
\newcommand{\fun}{f}
\newcommand{\responsedist}{\Psi}
\newcommand{\responsedens}{\psi}
\newcommand{\redist}{N}
\newcommand{\re}{u}
\newcommand{\remle}{\widehat{\re}}
\newcommand{\cov}{x}
\newcommand{\basisfun}{b}
\newcommand{\splineweight}{\beta}
\newcommand{\splineweightmle}{\widehat{\splineweight}}

\newcommand{\resd}{\sigma}
\newcommand{\resdest}{\widehat{\resd}}
\newcommand{\betaest}{\widehat{\beta}}
\newcommand{\params}{\theta}
\newcommand{\paramsmle}{\widehat{\params}}

\newcommand{\smoothingparam}{\lambda}
\newcommand{\smoothingparamest}{\widehat{\smoothingparam}}

\newcommand{\penlik}{\mathcal{L}}
\newcommand{\penlikwithargs}{\penlik(\re, \splineweight;\response, \params)}
\newcommand{\penlikwithmaxargs}{\penlik\{\remle(\params), \splineweightmle(\params);\response, \params\}}
\newcommand{\marglik}{\mathcal{M}}
\newcommand{\marglikwithargs}{\marglik(\params;\response)}
\newcommand{\marglikapprox}{\widetilde{\marglik}}
\newcommand{\marglikapproxwithargs}[1]{\marglikapprox_{#1}(\params;\response)}
\newcommand{\hessian}{H}
\newcommand{\smoothpen}{\mathcal{P}}

\newcommand{\smoothingmat}{S}

\begin{document}

\title{Inference for generalized additive mixed models via penalized marginal likelihood}

\author{A. STRINGER \\Department of Statistics and Actuarial Science \\ University of Waterloo ,\\ Waterloo N2L 2G3, Ontario, Canada. \\
alex.stringer@uwaterloo.ca}

\maketitle
\tableofcontents

\begin{abstract}
The Laplace approximation is sometimes not sufficiently accurate for smoothing parameter estimation in generalized additive mixed models. 
A novel estimation strategy 
is proposed that solves this problem and leads to estimates exhibiting the correct statistical properties.
\end{abstract}

\section{Introduction}

A generalized additive mixed model for a response $\response = (\response_1\Tr,\ldots,\response_\numgroups\Tr)\Tr\in\Reals^{\totalsize}$ where each $\response_i = (\response_{i1},\ldots,\response_{i\groupsize_i})\Tr\in\Reals^{\groupsize_i}$ with $\totalsize = \groupsize_1 + \cdots + \groupsize_\numgroups$ is:
\begin{equation}\label{eqn:gammdefinition}
\response_{ij} | \re_i \indsim \responsedist\{\mean(\cov_{ij}, \re_i)\}, \ \link\{\mean(\cov,\re)\} = \sum_{s=1}^{\numfun}\fun_s(\cov_s) + \re, \ \re_i \iidsim \redist\left(0, \resd^2\right).
\end{equation}
Here $\cov_{ij}\in\Reals^{\covdim}$ for $i=1,\ldots,\numgroups, j=1,\ldots,\groupsize_i$, $\re_i\in\Reals$, $\responsedist$ is a distribution having suitably smooth density $\responsedens$,
$\mean(x,u) = E(y|x, u)$, and $g$ is a link function.
The random effects $\re_i,i=1,\ldots,\numgroups$ induce dependence between observations $\response_{ij}, \response_{il}$ in the same group.
The unknown smooth functions $\fun_1,\ldots,\fun_\numfun$ allow the mean to depend on the covariates in a nonlinear manner and are represented by the basis expansions
$
\fun_s(x) = \basisfun_{1s}(x)\splineweight_{1s} + \cdots + \basisfun_{\fundim_{s}s}(x)\splineweight_{\fundim_{s}s},
$
where $\basisfun_{ls}$ is the $l$th cubic B-spline basis function for function $\fun_s$ on a knot sequence of appropriate length, and $\splineweight_{ls}$ is the corresponding spline weight to be estimated from the data.
The full vector of $\fundim = \fundim_1 + \cdots + \fundim_\numfun$ unknown spline weights is $\splineweight = (\splineweight_1\Tr,\ldots,\splineweight_\numfun\Tr)\Tr\in\Reals^{\fundim}$
where $\splineweight_s = (\splineweight_{s1},\ldots,\splineweight_{s\fundim_s})\Tr$.
Estimation of $\splineweight$ and prediction of $\re$ is based on minimizing the negative penalized log-likelihood,
\begin{align}\label{eqn:penalizedlikelihood}
&\penlikwithargs = \sum_{i=1}^{\numgroups}\penlik_i(\re_i,\beta;\response,\params) + \smoothpen(\beta, \lambda), \\
&= -\sum_{i=1}^{\numgroups}\left[\sum_{j=1}^{\groupsize_i}\log\responsedens\{\response_{ij};\mean(x_{ij}, \re_i)\} + \log g(u_i;\sigma)\right]
+ \frac{1}{2}\sum_{s=1}^{\numfun}\left[\smoothingparam_s\int\left\{\fun^{\prime\prime}_s(x)\right\}^2dx - (\fundim_s-2)\log\lambda_s\right],
\end{align}
\citep{wood_fast_2011,wood_straightforward_2013}, where $\params = (\resd, \smoothingparam_1,\ldots,\smoothingparam_\numfun)\Tr\in\Reals^{\numfun+1}$ contains the random effect variance and smoothing penalty parameters, both of which must be estimated, and $g(u;\sigma)$ is the density of $u\sim\text{N}(0,\sigma^2)$.
A connection between penalized smoothing and random effects models is observed by writing the penalty as a quadratic form in $\splineweight$,
$$
\int\left\{\fun^{\prime\prime}_s(x)\right\}^2dx = \splineweight_s\Tr\smoothingmat_s\splineweight_s, \ \left(\smoothingmat_s\right)_{lr} = \int\basisfun^{\prime\prime}_{ls}(x)\basisfun^{\prime\prime}_{rs}(x)dx, \
\text{dim}(S_s) = \fundim_s, \ \text{rank}(S_s) = \fundim_s-2,
$$
and hence interpreting it as an improper Gaussian prior on $\splineweight$ with precision matrix $\smoothingmat_\smoothingparam = \text{blockdiag}(\smoothingparam_1\smoothingmat_1,\ldots,\smoothingparam_\numfun\smoothingmat_\numfun)$.
It follows that $\exp\{-\smoothpen(\beta, \lambda)\}$ is proportional to a (low rank) Gaussian density with precision matrix $\smoothingmat_\smoothingparam$, and
hence $\penlik$ is a negative joint log-likelihood of $\re, \splineweight$.
Interpreting the penalized smooths as random effects leads to inference for $\params$ based on minimizing the negative marginal log-likelihood \citep{wood_fast_2011},
\begin{equation}\label{eqn:marginallikelihood}
\marglikwithargs = -\log\int\exp\left\{-\penlikwithargs\right\}d\re d\splineweight,
\end{equation}
so $\paramsmle = \text{argmin }\marglikwithargs$.
However, when $\responsedist$ is not a Gaussian distribution the integral (\ref{eqn:marginallikelihood}) is intractable and $\paramsmle$ cannot be calculated.
Instead, inference is based on minimizing some approximation to $\marglik$. 
Current methods in the literature \citep{wood_straightforward_2013} and in software (package \texttt{mgcv}, \citealt{wood_fast_2011} and package \texttt{gamm4}, \citealt{wood_gamm4_2020})
employ the Laplace approximation for this purpose,
\begin{equation}\label{eqn:laml}
\marglikapproxwithargs{LA} = -\frac{\numgroups + \fundim}{2}\log(2\pi) + \frac{1}{2}\log\det\{\hessian(\params)\} - \penlikwithmaxargs,
\end{equation}
where $\remle(\params) = \text{argmin}_\re\penlikwithargs$, $\splineweightmle(\params) = \text{argmin}_\splineweight\penlikwithargs$, and $\hessian(\params)$ is the Hessian of $\penlik$ with respect to $\re,\splineweight$ at $\remle(\params),\splineweightmle(\params)$ for given $\params$.
While the Laplace approximation is known to be acceptable for smoothing penalty parameter estimation in spline models without group-specific random effects \citep{kauermann2009some}, it is often not sufficiently
accurate for variance component estimation in generalized linear mixed models with group-specific random effects \citep{joe_accuracy_2008,stringer2023approximate,bilodeau2024asymptotics}.
The use of the Laplace approximation for variance in generalized additive mixed models has not been directly investigated, but these analyses for the linear case suggest that it may not be appropriate.
To see the potential problem, observe that the marginal likelihood factors in the following manner:
\begin{equation}\label{eqn:factormarglik}
\exp\left\{-\marglikwithargs\right\} = \int\left[\prod_{i=1}^{\numgroups}\int\exp\left\{-\penlik_i(\re_i,\beta;\response, \params)\right\}d\re_i\right]\exp\left\{-\smoothpen(\beta,\lambda)\right\} d\splineweight.
\end{equation}
Inspection of (\ref{eqn:factormarglik}) reveals that the $d\re$ integral in (\ref{eqn:marginallikelihood}) factors over $\re_1,\ldots,\re_\numgroups$ due to their assumed independence.
However, $\numgroups\to\infty$ is required for consistency of $\paramsmle$, which is in turn required for consistency of $\splineweightmle$.
For generalized linear mixed models \citet{ogden_asymptotic_2017} gives a thorough analysis of lower bounds and argues that $\numgroups$ cannot grow too fast compared to $\groupsize_{\text{min}}$ if consistent estimates are desired.
Figure \ref{fig:covrplot} in section \ref{sec:simulations} shows the coverage of confidence intervals for $f$ decreasing as $m$ is increased in a simulated example, illustrating the practical failure of the Laplace approximation. 
In contrast, adaptive Gaussian quadrature is widely recognized as an appropriately accurate method for marginal likelihood approximation in mixed models \citep{pinheiro1995approximations,pinheiro2006efficient}.
\citet{bilodeau2024asymptotics,stringer2023approximate} show futher simulations and provide a stochastic upper bound on the error in using adaptive Gaussian quadrature to fit generalized linear mixed models which shows that using this more accurate integral approximation mitigates the problem. Unfortunately, in generalized additive mixed models, while adaptive quadrature could be applied to each one-dimensional $du_i$ integral,
the dimension of the $d\beta$ integral is too large for this technique to be computationally feasible.

\section{Two-stage inference in generalized additive mixed models}\label{sec:newapproach}

We propose to ignore dependence between the observations for the purposes of smoothing parameter estimation, which we address using a standard generalized additive model
fit by Laplace-approximate marginal likelihood or restricted marginal likelihood.
We then propose to fit a generalized linear mixed model with a fixed penalty for $\beta$, using the estimated smoothing parameters from the first step. 
This follows the common practice of ignoring uncertainty in the estimation of $\lambda_1,\ldots,\lambda_\numfun$, but adequately captures the uncertainty in $\beta$ and hence $f$, leading
to confidence intervals whose coverages appear to attain the nominal level as $m\to\infty$.

First consider the generalized additive model,
\begin{equation}\label{eqn:gam}
\response_{ij} \indsim \responsedist\{\mean(\cov_{ij})\}, \ \link\{\mean(\cov)\} = \sum_{s=1}^{\numfun}\fun_s(\cov_s),
\end{equation}
which is Eq. (\ref{eqn:gammdefinition}) with $\sigma = 0$. This model is fit by employing the Laplace-approximate marginal or restricted marginal likelihood method of \citet{wood_fast_2011} through the \texttt{mgcv} package.
Let $\smoothingparamest = (\smoothingparamest_1,\ldots,\smoothingparamest_\numfun)\Tr$ be the estimated smoothing parameters obtained in this manner,
and let $\smoothingmat_{\smoothingparamest} = \text{blockdiag}(\smoothingparamest_1\smoothingmat_1,\ldots,\smoothingparamest_\numfun\smoothingmat_\numfun)$.
Let $z\in Q(k)\subset\Reals^{+}\cup\{0\}$ be the points and $w_k:Q(k)\to\Reals^{+}$ the weights from a Gauss-Hermite quadrature rule of order $k$ \citep[Eqs. 3 and 4]{bilodeau2024stochastic},
$\remle_i\equiv \remle_i(\sigma, \beta) = \text{argmin}_\re\penlik_i(\re, \beta;\response, \sigma, \smoothingparamest)$,
and $h_{ii}\equiv h_{ii}(\sigma, \beta) = \partial^2_{\re^2}\penlik_i\{\remle_i(\sigma, \beta), \beta;\response, \sigma\}$.
We propose the following penalized approximate log-marginal likelihood for estimating $\sigma,\beta$:
\begin{equation}\label{eqn:penmarglik}
\exp\left\{-\marglikapprox_{k}(\sigma, \beta)\right\} = 
\left[\prod_{i=1}^{m}h_{ii}^{-1/2}\sum_{z\in Q(k)}w_{k}(z)\penlik_i\{\remle_i + z\cdot h_{ii}^{-1/2}, \beta;\response,\sigma\}\right]
\exp\left\{-\smoothpen(\beta, \smoothingparamest)\right\}.
\end{equation}
We estimate $(\resdest, \betaest)\Tr = \text{argmin }\marglikapprox_{k}(\sigma, \beta)$ and form Wald confidence intervals for $\beta$ using standard errors obtained from the 
diagonal elements of the inverse Hessian of $\marglikapprox_{k}(\resdest, \betaest)$.

The accuracy of the approximation (\ref{eqn:penmarglik}) is determined by the order of the quadrature rule, $k\in\Nats$.
In generalized linear and non-linear mixed models,
\citet{bilodeau2024asymptotics} show that under assumptions on the model that include the exponential family, for any $\varepsilon>0$ if $m = n_{min}^q$ for some $q>0$ then the relative approximation error is $O_p[R_n n_{min}^{-\{r(k)+1\}/2 + \varepsilon}]$ where $r(k) = \lfloor (k+2)/3 \rfloor$ and $R_n = m^{1/2}$ or $(mn)^{1/2}$ is the parameter-dependent rate of convergence of the maximum likelihood estimator based on the exact marginal likelihood \citep{jiang2022usable}. 
For fixed $\beta$ this result is expected to apply here without modification, since all that is changed is the addition of the 
$\exp\{-\smoothpen(\beta,\smoothingparamest)\}$ term which does not depend on $m,n$. 
The practical implication, as discussed by \citet{bilodeau2024asymptotics}, is that $k$ can always be chosen high enough for a given set of data such that the sampling error in $\response$
dominates the numerical error in the integral approximation, rendering inferences indistinguishable from those that would be obtained if the exact marginal likelihood could be calculated.

\section{Empirical Analysis}\label{sec:simulations}

A simulation study was conducted to (a) illustrate the inadequacy of Eq. (\ref{eqn:laml}) for inferences in the generalized additive mixed model (\ref{eqn:gammdefinition}),
and (b) provide empirical evidence of the adequacy of the proposed penalized marginal likelihood given by Eq. (\ref{eqn:penmarglik}) for these inferences.
Code for reproducing these results is available at https://github.com/awstringer1/gamm-paper-code.
The Laplace-approximate generalized additive mixed model was fit using the \textsf{R} package \texttt{gamm4} \citep{wood_gamm4_2020} and the generalized additive model is fit using the R package \textsf{mgcv} \citep{wood_fast_2011}.
The \texttt{pml} method is implemented in \textsf{R} package
\texttt{aghqmm} \citep{stringer2023approximate}.
More extensive simulation results covering multiple smooth functions, more wiggly and more flat functions, varying $\sigma$ and $k$, and smaller $m$ are presented in the supplementary materials to this paper. 
The extreme case shown in supplement section A.10 with $f(x) = 2x-1$ lying in the penalty nullspace
was observed to yield \texttt{pml} estimates with average
coverage too low---comparable to the \texttt{gamm}---presumably
due to the difficulty of estimating $\lambda$ when $f$
is linear. However, the bias of $f$ and $\sigma$ was still zero on average with \texttt{pml}, and the latter nonzero for \texttt{gamm}.
In the case that $f$ is estimated to be linear, neither the \texttt{pml} nor the \texttt{gamm} methods should be used to make confidence intervals for $f$.
All other cases are consistent with the results presented here.

\begin{figure}
\centering
\begin{subfigure}{0.7\textwidth}
	\includegraphics[width = 4in, height = 4in]{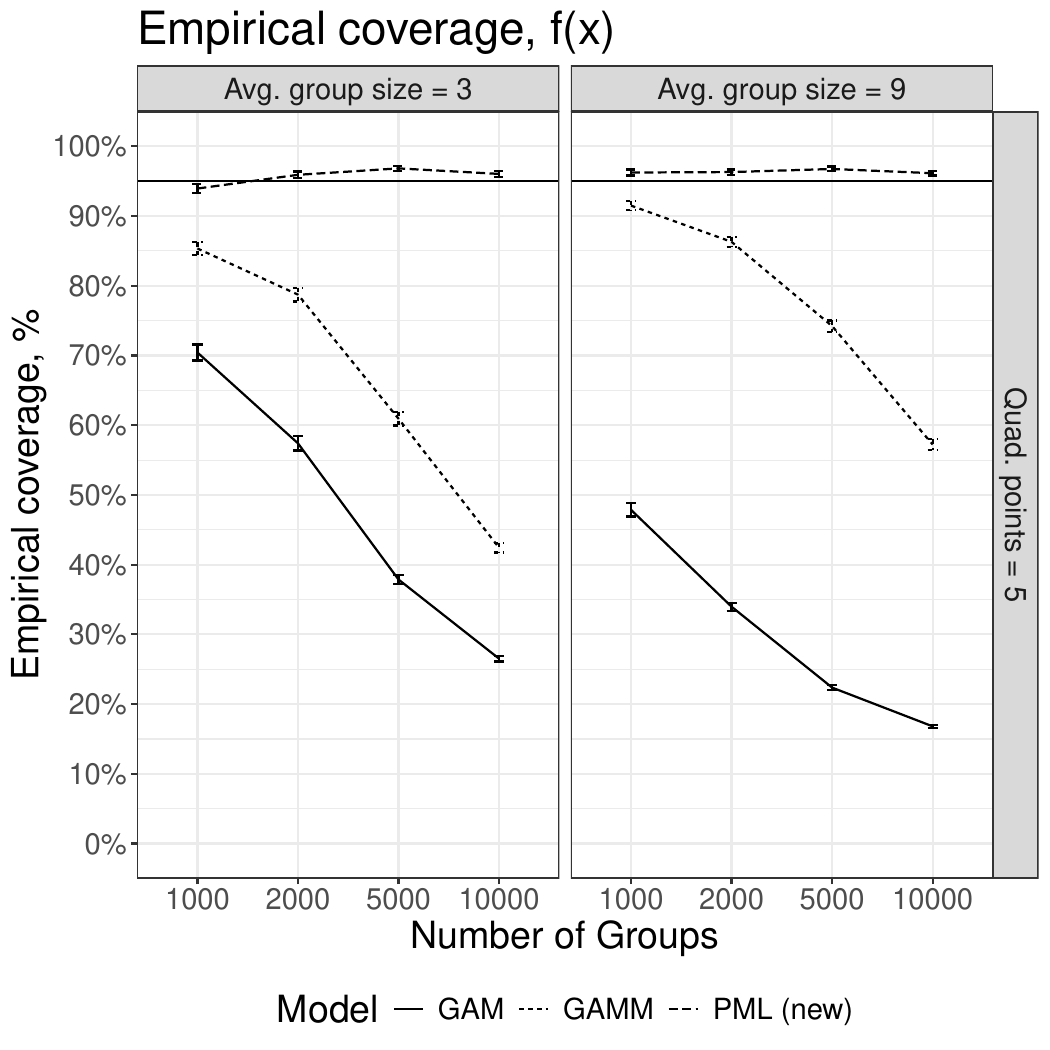}
	\caption{Empirical coverage in estimated $f(x)$ across $1,000$ simulations. 
	}
	\label{fig:covrplot}
\end{subfigure}

\begin{subfigure}{0.7\textwidth}
	\includegraphics[width = 4in, height = 4in]{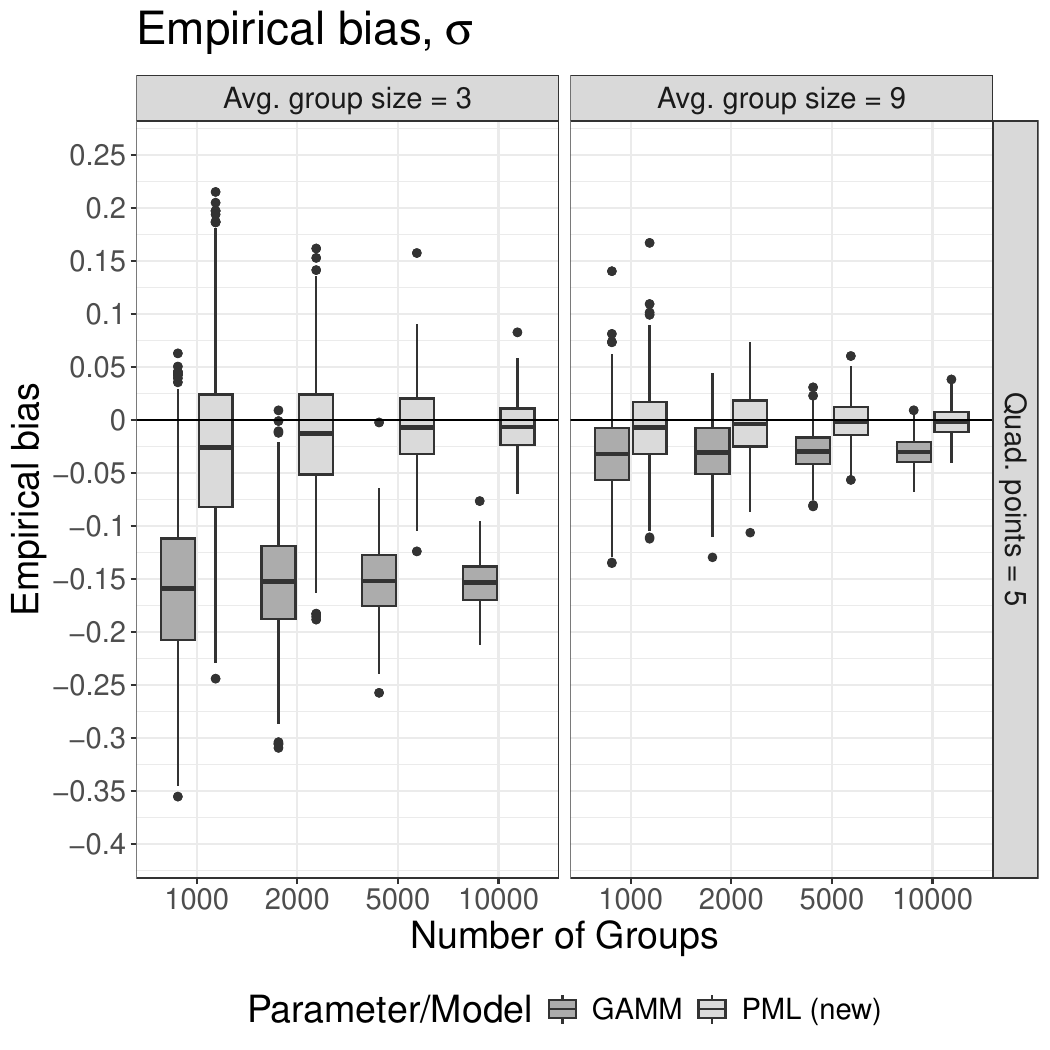}
	\caption{Empirical bias in estimated $\sigma$ across $1,000$ simulations. 
	}
	\label{fig:biasplotsigma}
\end{subfigure}
\caption{Simulation results, coverage of $\widehat{f}$ and bias of $\widehat{\sigma}$ from the generalized additive model,
generalized additive mixed model, and penalized marginal likelihood.}
\label{fig:results}
\end{figure}

Replicate sets of data with equal-sized groups were generated from model (\ref{eqn:gammdefinition}) with $\numfun = 1, f(x) = \sin(2\pi x), \sigma = 1$, varying $m, n$, and Bernoulli response. The covariates $x_{ij}$ are generated independently from a $\text{U}(0,1)$ distribution.
To each of the simulated data sets, the model (\ref{eqn:gammdefinition}) was fit using (a) a standard generalized additive model that ignores dependence in $\response$ (GAM), (b) the existing method by minimizing $\marglikapproxwithargs{LA}$ (GAMM), and (c) the new penalized marginal likelihood method described in section \ref{sec:newapproach}, with $k = 5$ quadrature points (PML).
The bias in $f$ and $\sigma$ are reported and the average across-the-function coverage in $f$ is compared to the nominal value of $95\%$.
The results shown in Figure \ref{fig:results} are based on $1,000$ simulated sets of data.

All three approaches are successful at point estimation of $f$, having zero average empirical bias
for all values of $m$ and $n$ considered (Figure 1 in the supplement).
Figure \ref{fig:covrplot} shows empirical coverages. The generalized additive model has empirical coverage that drops substantially as both $m$ 
and $n$ are increased, which is expected as this model incorrectly ignores dependence in $\response$.
The Laplace-approximate generalized additive mixed model captures dependence in $\response$, but still has
empirical coverage that decreases as $m$ is increased. 
In contrast to the generalized additive model,
this behaviour is less severe when $n$ is larger. The explanation is that this behaviour
is due to the inadequacy of the Laplace approximation, a result consistent with \citet{joe_accuracy_2008,stringer2023approximate,bilodeau2024asymptotics}.
In contrast, the penalized marginal likelihood
approach exhibits nominal average coverage for all values of $m$ and $n$ tried, since it relies
on an appropriately accurate approximation to the marginal likelihood and incorporates appropriate
penalization of $\beta$.
Figure \ref{fig:biasplotsigma} shows empirical bias for estimation of $\sigma$ from the generalized additive mixed model and penalized marginal likelihood. The generalized additive model sets $\sigma=0$ so does not return an estimate. The generalized additive mixed model shows bias converging to a nonzero value as $m$ is increased for both values of $n$, with the effect less
severe for larger $n$. The penalized marginal likelihood again corrects this behaviour with
empirical bias converging to zero as $m$ is increased, and smaller bias for larger $n$.

\section*{Acknowledgements}
This work was funded by NSERC grant RGPIN-2023-03331.

\bibliographystyle{chicago}
\bibliography{references}


\newpage

\begin{center}
{\huge Supplementary materials for:\\
Inference for generalized additive mixed models via penalized marginal likelihood}
\end{center}

\newcommand{\simdate}{20250306} 
\newcommand{\simdateflat}{20250409} 

\newcommand{\plotwidth}{6in}
\newcommand{\plotheight}{\plotwidth}

\renewcommand*{\thesection}{\Alph{section}}
\setcounter{section}{0}

\section{Simulations}

\subsection{Reproduction}

The code required to reproduce this simulation study is found on github at \href{https://github.com/awstringer1/gamm-paper-code}{https://github.com/awstringer1/gamm-paper-code}.

\subsection{Setup}

Simulations are performed based on model (1) from the main manuscript:
\begin{equation}\label{eqn:gammdefinition}
\response_{ij} | \re_i \indsim \responsedist\{\mean(\cov_{ij}, \re_i)\}, \ \link\{\mean(\cov,\re)\} = \sum_{s=1}^{\numfun}\fun_s(\cov_s) + \re, \ \re_i \iidsim \redist\left(0, \resd^2\right).
\end{equation}
Here $i = 1,\ldots,m$ indexes subjects/groups and $j = 1,\ldots,n_i$ indexes
observations/measurements on each group.
For each simulation study $B\in\Nats$ replicate sets of data are generated from
model (\ref{eqn:gammdefinition}) for a given $m$ and $n_1,\ldots,n_m$, with $x_{ij}$ generated independently from a $\text{U}(0,1)$ distribution.
Group sizes $n_i$ are unbalanced and chosen such that they are equal to a value $n$ on average, and this $n$ is reported in each simulation.
Estimates $\widehat{f}_b$ and $\widehat{\sigma}_b$ are returned for each set $b = 1,\ldots,B$ of simulated data.
A pointwise confidence interval $\widehat{C}_{\alpha}(x; \widehat{f}), \alpha = 0.05$ is returned for any $x\in\mathbb{R}$, and should satisfy $P\{f(x)\in\widehat{C}_{\alpha}(x; \widehat{f})\}\approx\alpha$ for any $x$ where $P$ is the distribution of the data implied by (\ref{eqn:gammdefinition}).
For a fine grid $x_1,\ldots,x_N$ of some large size $N\in\Nats$, it is then expected that the average-average-across-the-function coverage is close to nominal, that is $N^{-1}\sum_{i=1}^{N}I\{f(x_i)\in\widehat{C}_{\alpha}(x_i; \widehat{f}_b(x_i))\}\approx\alpha$.

Performance is measured as follows. For a fine grid $x_1,\ldots,x_N$, $N=1000$,
compute:
\begin{enumerate}
  \item \textbf{Bias} of $\widehat{f}$: boxplot over $b=1,\ldots,B$ of $N^{-1}\sum_{i=1}^{N}\{\widehat{f}_b(x_i) - f(x_i)\}$.
  \item \textbf{Bias} of $\widehat{\sigma}$: boxplot over $b=1,\ldots,B$ of $\widehat{\sigma}_b - \sigma$.
  \item \textbf{Coverage} of $\widehat{f}$: plot $\widehat{E} = B^{-1}N^{-1}\sum_{b=1}^{B}\sum_{i=1}^{N}I\{f(x_i)\in\widehat{C}_{0.05}(x_i; \widehat{f}_b(x_i))\}$ as points connected by lines with $m$ on the $x$-axis, with error bars given by $\Phi^{-1}(.975)$-times the Monte Carlo standard error of $\sqrt{\widehat{E}(1-\widehat{E})/B}$.
\end{enumerate}
For each simulation setup, results from three methods are reported:
\begin{enumerate}
  \item \texttt{gam}: a generalized additive model fit using the \texttt{mgcv::bam} function using options \texttt{method = "REML"} and \texttt{discrete = FALSE},
  \item \texttt{gamm}: a generalized additive mixed model (Laplace approximation) fit using the \texttt{gamm4::gamm4} function with options \texttt{REML = TRUE},
  \item \texttt{pml}: the new penalized marginal likelihood method with $k\in\mathbb{N}$ quadrature points.
\end{enumerate}
Specific choices of $p, \sigma, f$ and $k$ are given in the individual subsections.

\subsection{Main manuscript}

\textbf{Setup}:
\begin{itemize}
  \item $B = 1000$.
  \item $\responsedist(\mean) = \text{Bernoulli}(\mean), \link(\mean) = \log(\mean/(1-\mean))$.
  \item $m = 1000, 2000, 5000, 10000$.
  \item $n_1,\ldots,n_m$: computed using \texttt{sample(2:(2 * (n - 1)), size = m, replace = TRUE)} with $n = 3, 9$.
  \item $\sigma_u = 1$.
  \item $k = 5$.
  \item $p = 1, f(x) = \sin(2\pi x)$.
\end{itemize}

\begin{figure}[p]
  \centering
  \includegraphics[width = \plotwidth, height = \plotheight]{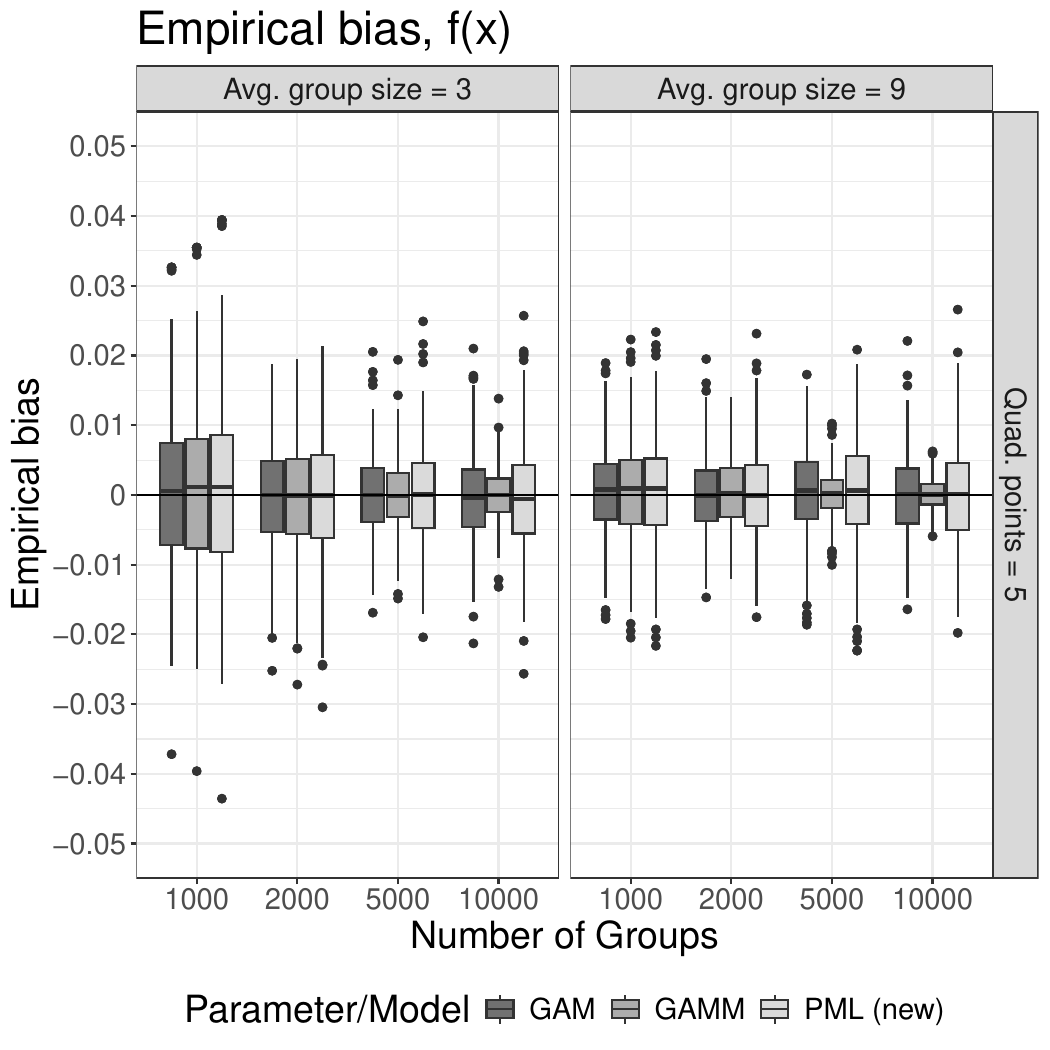}
  \caption{Empirical bias of $\widehat{f}$, true $f(x) = \sin(2\pi x)$, varying $m$ and $n$.}
  \label{fig:mainbias}
\end{figure}

\begin{figure}[p]
  \centering
  \includegraphics[width = \plotwidth, height = \plotheight]{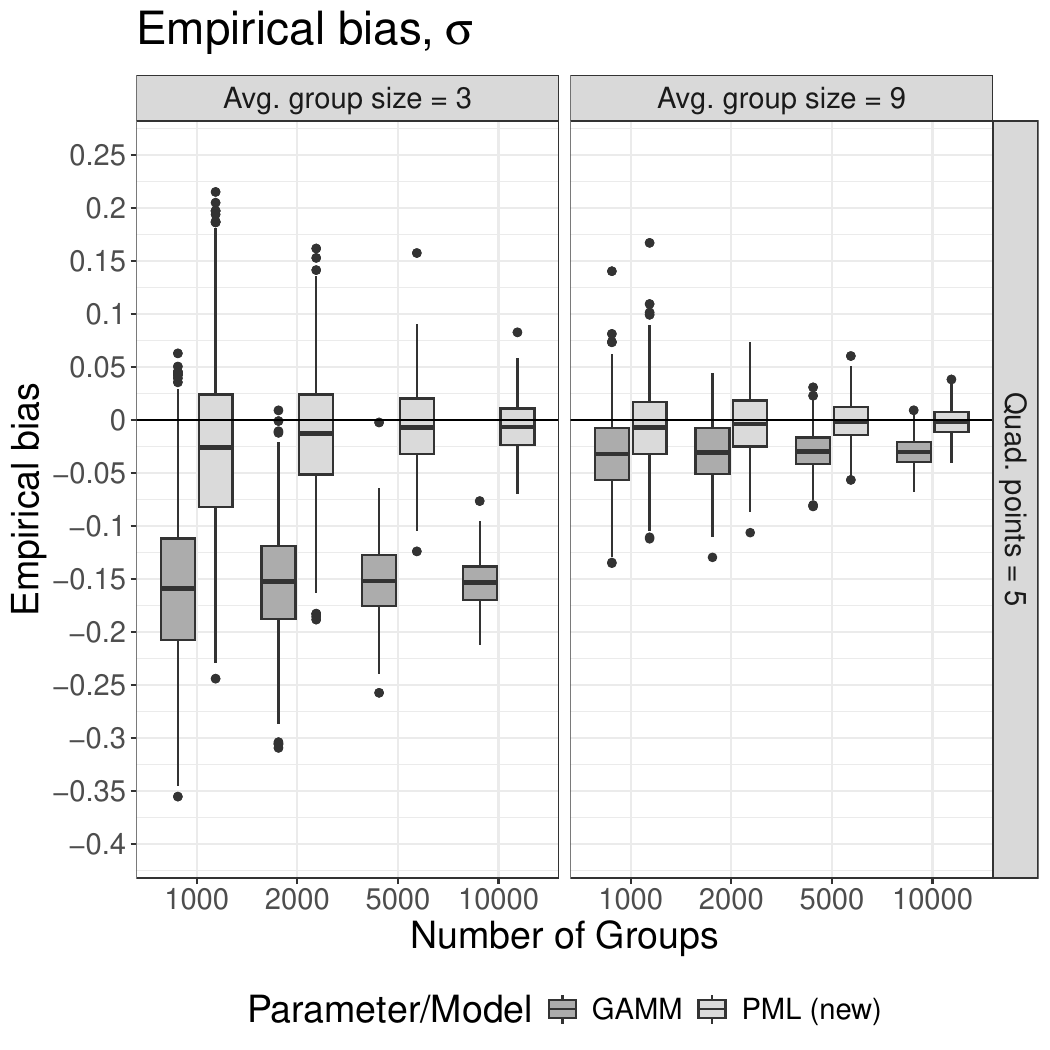}
  \caption{Empirical bias of $\widehat{\sigma}$, true $f(x) = \sin(2\pi x)$, varying $m$ and $n$.}
  \label{fig:mainsigma}
\end{figure}

\begin{figure}[p]
  \centering
  \includegraphics[width = \plotwidth, height = \plotheight]{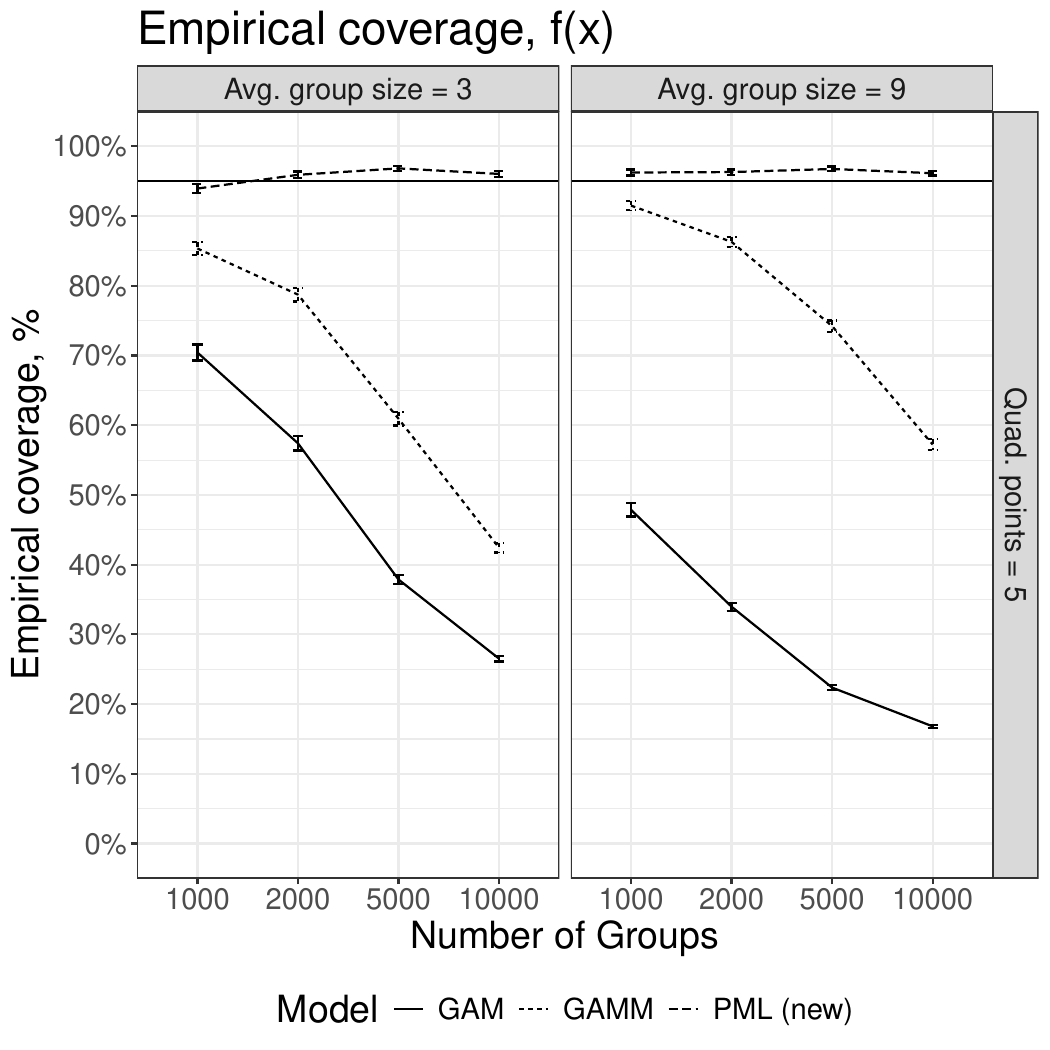}
  \caption{Empirical coverage of $\widehat{f}$, true $f(x) = \sin(2\pi x)$, varying $m$ and $n$.}
  \label{fig:maincovr}
\end{figure}

\textbf{Results}: all three methods have comparable average across-the-function bias of $\widehat{f}$, which is zero on average across the simulations (Figure \ref{fig:mainbias}).
The \texttt{gamm} shows non-zero average bias for $\widehat{\sigma}$ which appears to be converging to a nonzero value as $m$ is increased, an effect which is less severe for higher $n=9$ compared to lower $n=3$; the new \texttt{pml} method appears to have average bias for $\widehat{\sigma}$ converging to $0$ as $m$ is increased for both values of $n$ (Figure \ref{fig:mainsigma}).
The coverage of $\widehat{f}$ for the \texttt{gamm} decreases to far below the nominal level as $m$ is increased, while for \texttt{pml} it appears to level off at a slightly conservative value as $m$ is increased (Figure \ref{fig:maincovr}).

\subsection{Multiple smooth functions}

\textbf{Setup}:
\begin{itemize}
  \item $B = 500$
  \item $\responsedist(\mean) = \text{Bernoulli}(\mean), \link(\mean) = \log(\mean/(1-\mean))$.
  \item $m = 1000, 2000, 5000, 10000$.
  \item $n_1,\ldots,n_m$: computed using \texttt{sample(2:(2 * (n - 1)), size = m, replace = TRUE)} with $n = 3$.
  \item $\sigma_u = 1$.
  \item $k = 5$.
  \item $p = 2, f_1(x) = \sin(2\pi x), f_2(x) = \cos(2\pi x)$.
\end{itemize}

\begin{figure}[p]
  \centering
  \includegraphics[width = \plotwidth, height = \plotheight]{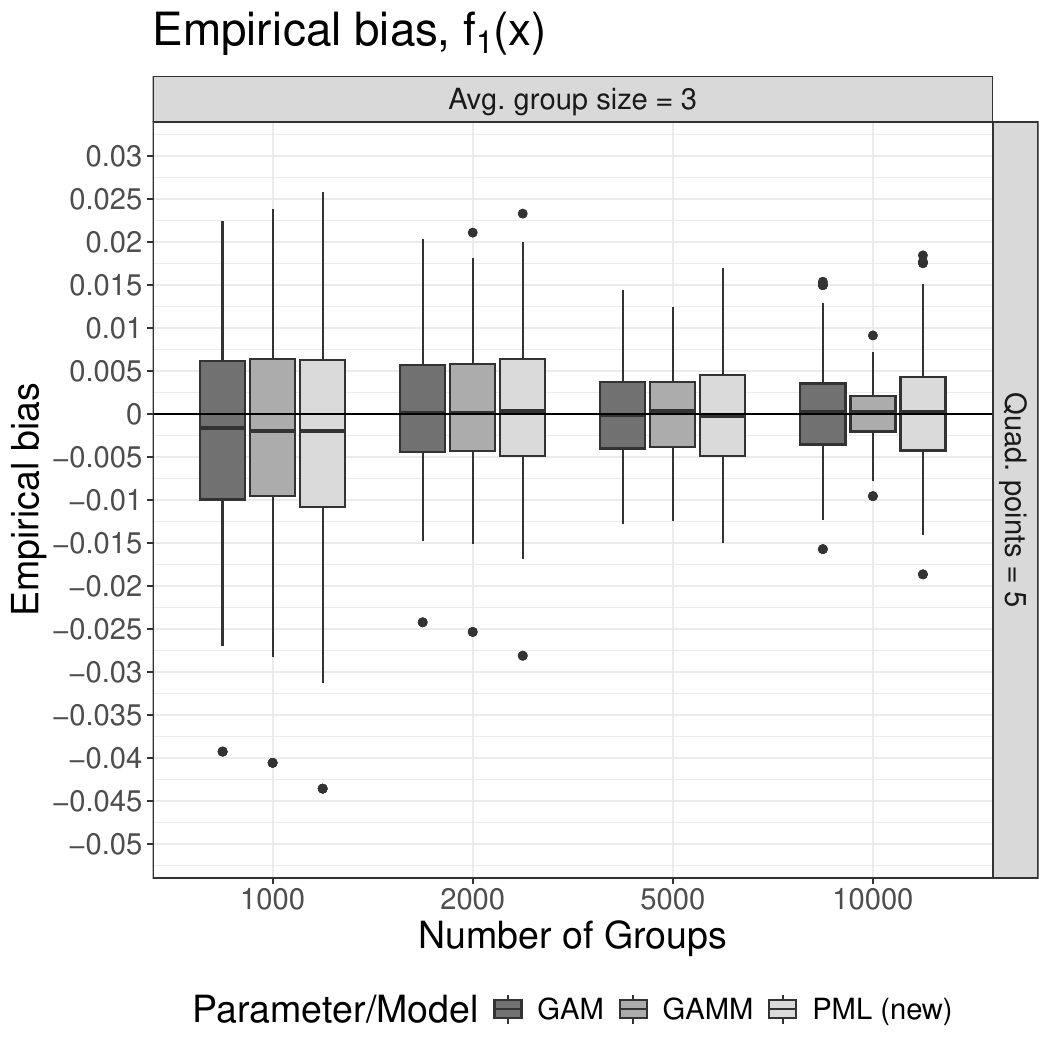}
  \caption{Empirical bias of $\widehat{f}_1$, true $f_1(x) = \sin(2\pi x)$, true $f_2(x) = \cos(2\pi x)$, varying $m$ and $n$.}
  \label{fig:multiplebias1}
\end{figure}

\begin{figure}[p][p]
  \centering
  \includegraphics[width = \plotwidth, height = \plotheight]{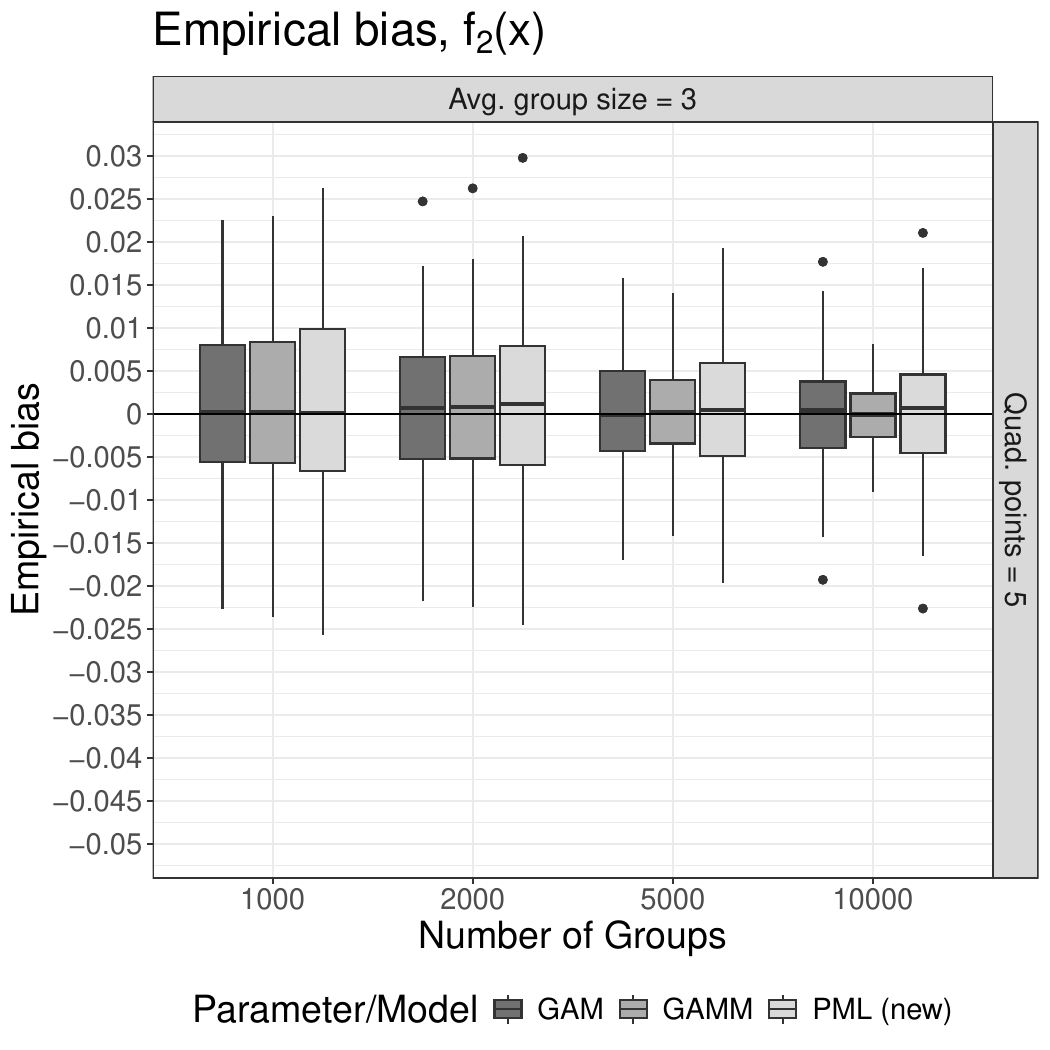}
  \caption{Empirical bias of $\widehat{f}_2$, true $f_1(x) = \sin(2\pi x)$, true $f_2(x) = \cos(2\pi x)$, varying $m$ and $n$.}
  \label{fig:multiplebias2}
\end{figure}

\begin{figure}[p][p]
  \centering
  \includegraphics[width = \plotwidth, height = \plotheight]{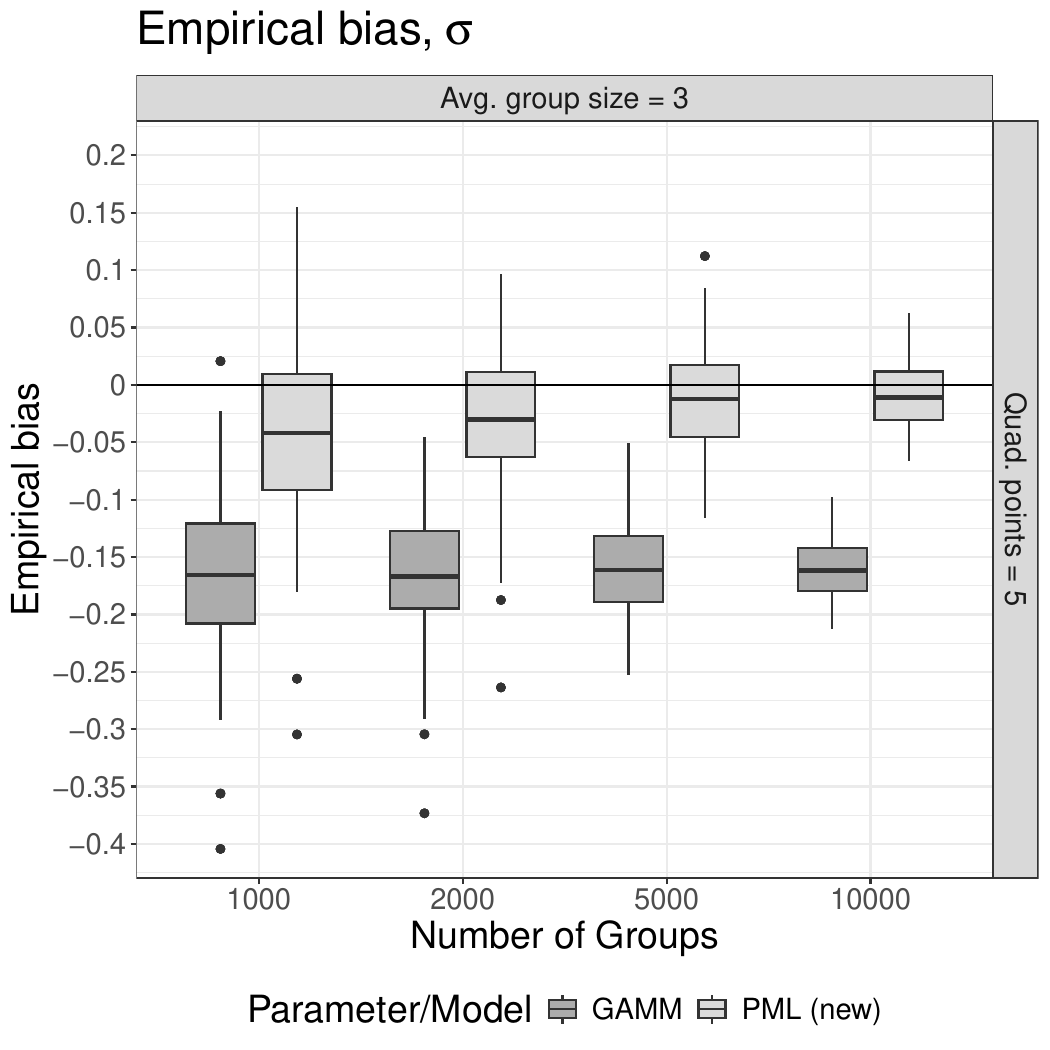}
  \caption{Empirical bias of $\widehat{\sigma}$, true $f_1(x) = \sin(2\pi x)$, true $f_2(x) = \cos(2\pi x)$, varying $m$ and $n$.}
  \label{fig:multiplesigma}
\end{figure}

\begin{figure}[p][p]
  \centering
  \includegraphics[width = \plotwidth, height = \plotheight]{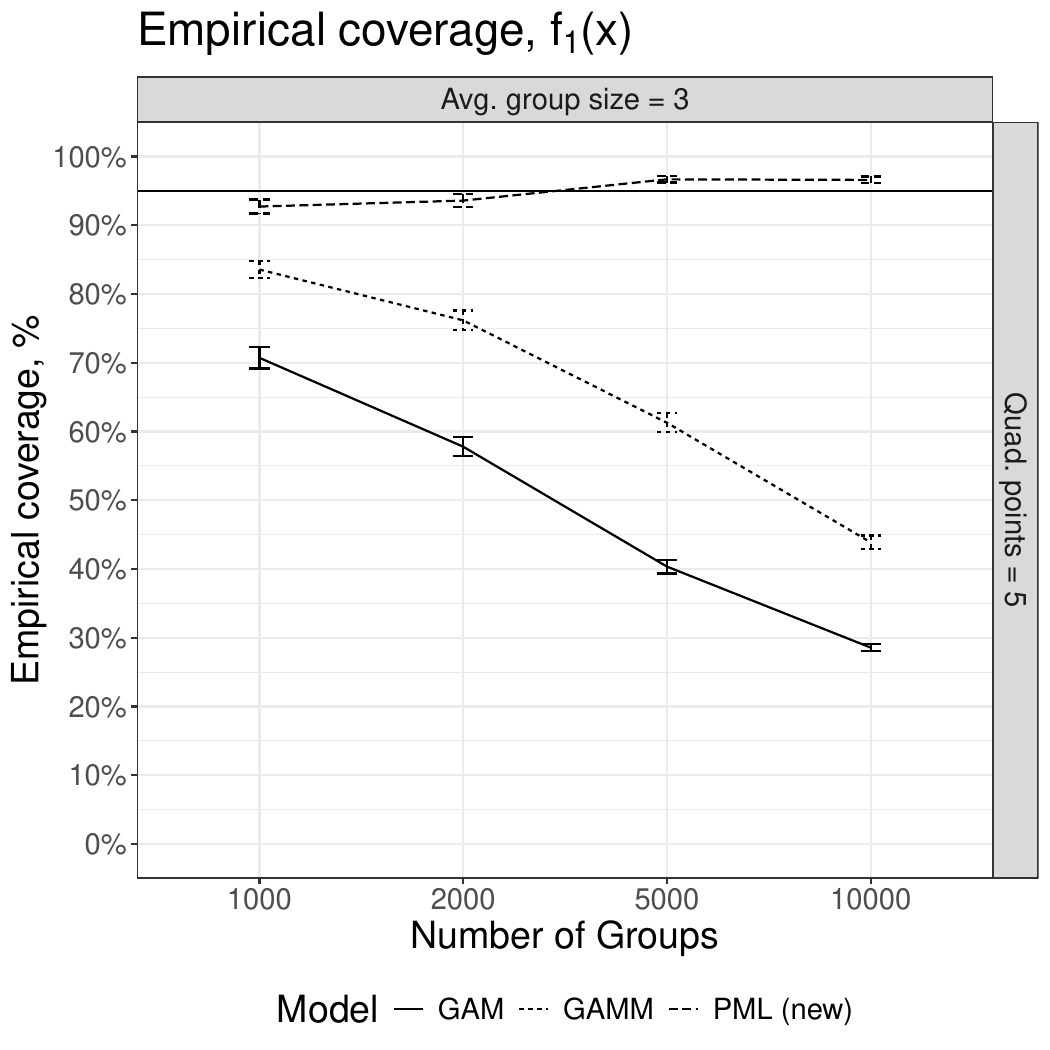}
  \caption{Empirical coverage of $\widehat{f}_1$, true $f_1(x) = \sin(2\pi x)$, true $f_2(x) = \cos(2\pi x)$, varying $m$ and $n$.}
  \label{fig:multiplecovr1}
\end{figure}

\begin{figure}[p][p]
  \centering
  \includegraphics[width = \plotwidth, height = \plotheight]{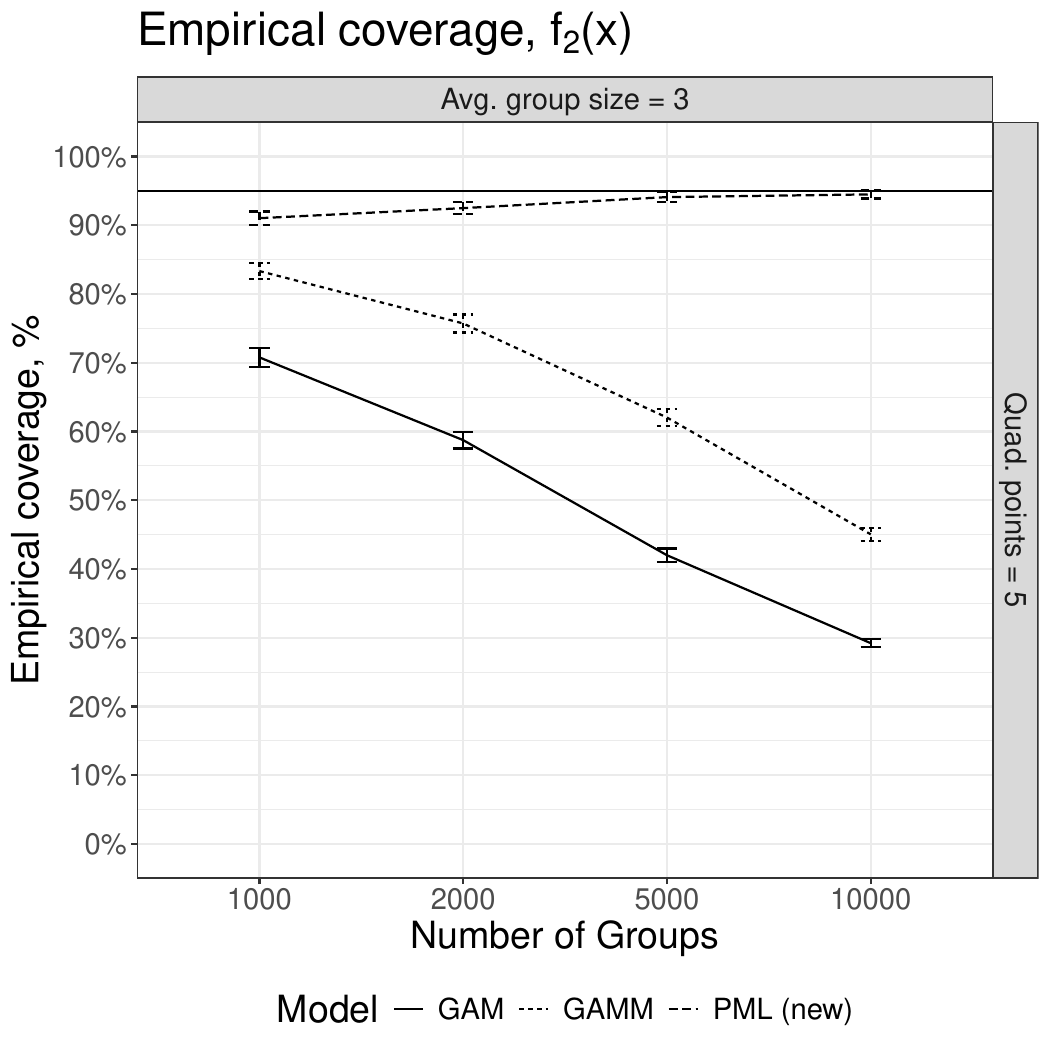}
  \caption{Empirical coverage of $\widehat{f}_2$, true $f_1(x) = \sin(2\pi x)$, true $f_2(x) = \cos(2\pi x)$, varying $m$ and $n$.}
  \label{fig:multiplecovr2}
\end{figure}

\textbf{Results}: all three methods have comparable average across-the-function bias of $\widehat{f}_1$ and $\widehat{f}_2$, which is zero on average across the simulations (Figure \ref{fig:multiplebias1} and \ref{fig:multiplebias2}).
The \texttt{gamm} shows non-zero average bias for $\widehat{\sigma}$ which appears to be converging to a nonzero value as $m$ is increased; the new \texttt{pml} method appears to have average bias for $\widehat{\sigma}$ converging to $0$ as $m$ is increased (Figure \ref{fig:multiplesigma}).
The coverage of both $\widehat{f}_1$ and $\widehat{f}_2$ for the \texttt{gamm} decreases to far below the nominal level as $m$ is increased, while for \texttt{pml} it appears to level off close to the nominal value as $m$ is increased (Figure \ref{fig:multiplecovr1} and \ref{fig:multiplecovr2}).

\subsection{Varying $\sigma$}

\textbf{Setup}:
\begin{itemize}
  \item $B = 500$
  \item $\responsedist(\mean) = \text{Bernoulli}(\mean), \link(\mean) = \log(\mean/(1-\mean))$.
  \item $m = 1000, 2000, 5000, 10000$.
  \item $n_1,\ldots,n_m$: computed using \texttt{sample(2:(2 * (n - 1)), size = m, replace = TRUE)} with $n = 3$.
  \item $\sigma_u = 1, 1.5, 2$.
  \item $k = 5$.
  \item $p = 1, f(x) = \sin(2\pi x)$.
\end{itemize}

\begin{figure}[p]
  \centering
  \includegraphics[width = \plotwidth, height = \plotheight]{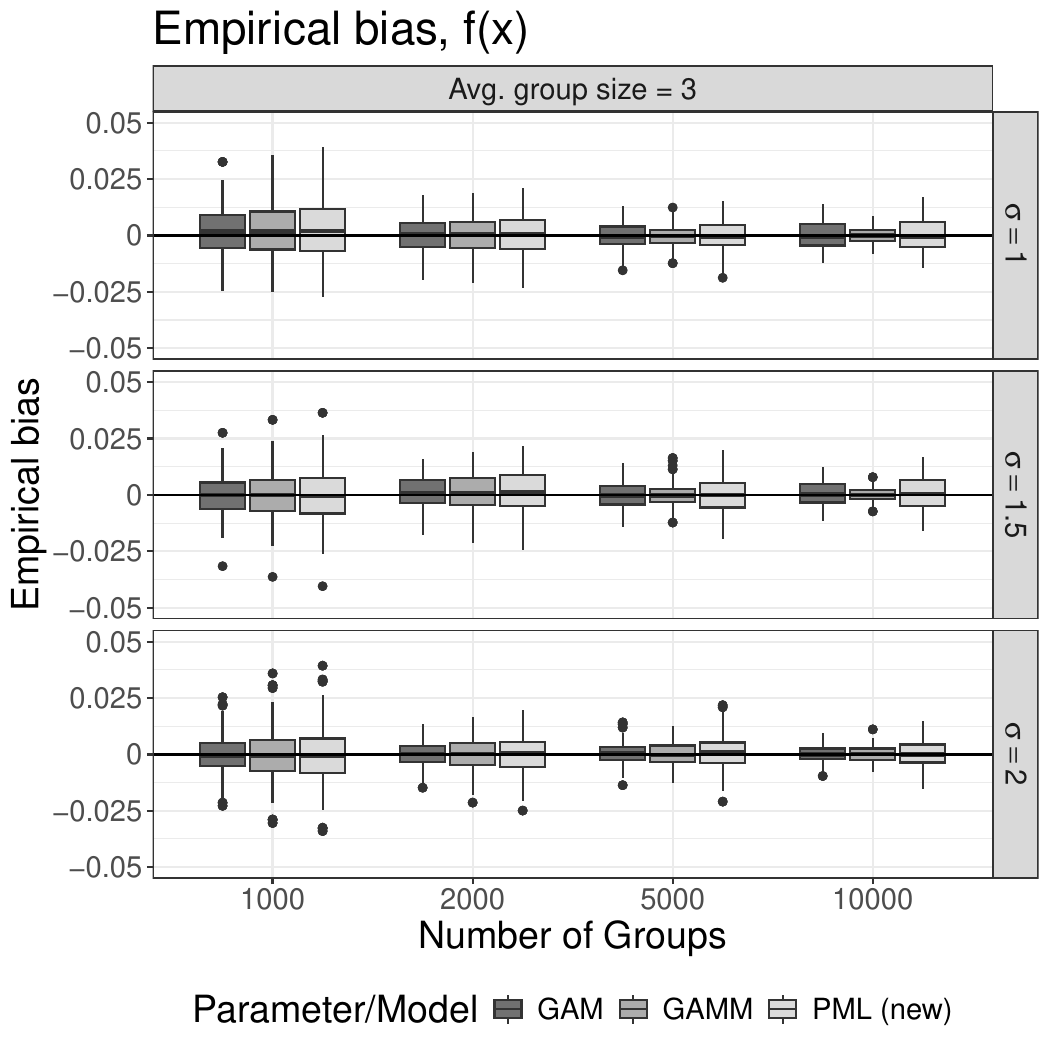}
  \caption{Empirical bias of $\widehat{f}$, true $f(x) = \sin(2\pi x)$, varying $m$ and $\sigma$.}
  \label{fig:sigmabias}
\end{figure}

\begin{figure}[p]
  \centering
  \includegraphics[width = \plotwidth, height = \plotheight]{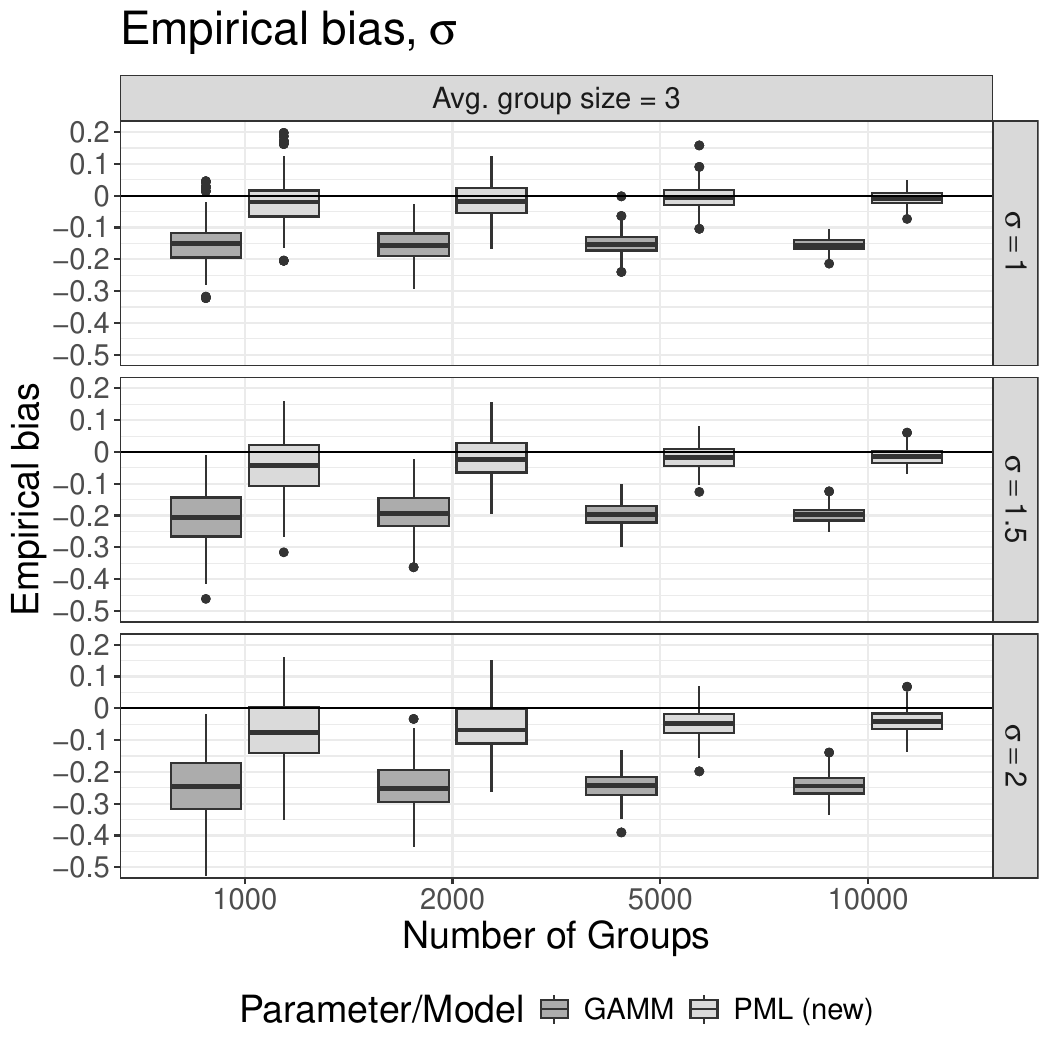}
  \caption{Empirical bias of $\widehat{\sigma}$, true $f(x) = \sin(2\pi x)$, varying $m$ and $\sigma$.}
  \label{fig:sigmasigma}
\end{figure}

\begin{figure}[p]
  \centering
  \includegraphics[width = \plotwidth, height = \plotheight]{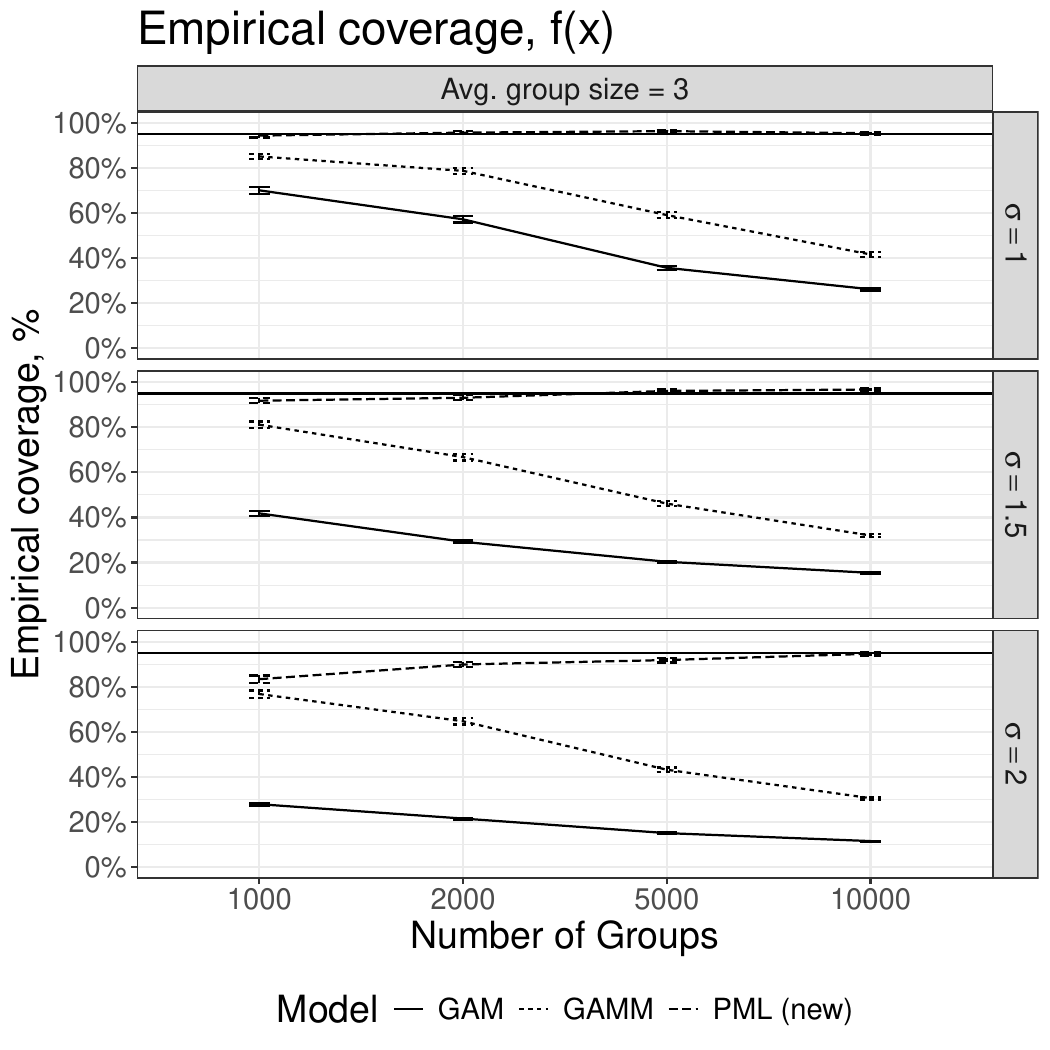}
  \caption{Empirical coverage of $\widehat{f}$, true $f(x) = \sin(2\pi x)$, varying $m$ and $\sigma$.}
  \label{fig:sigmacovr}
\end{figure}

\textbf{Results}: all three methods have comparable average across-the-function bias of $\widehat{f}$, which is zero on average across the simulations, for each value of $\sigma$ (Figure \ref{fig:sigmabias}).
The \texttt{gamm} shows non-zero average bias for $\widehat{\sigma}$ which appears to be converging to a nonzero value as $m$ is increased; the new \texttt{pml} method appears to have average bias for $\widehat{\sigma}$ converging to a value much closer to $0$ as $m$ is increased, however this value gets farther from $0$ for larger $\sigma$ (Figure \ref{fig:sigmasigma}).
The coverage of $\widehat{f}$ for the \texttt{gamm} decreases to far below the nominal level as $m$ is increased, while for \texttt{pml} it appears to level off at a slightly conservative value as $m$ is increased for all values of $\sigma$ (Figure \ref{fig:sigmacovr}).

\subsection{Varying $k$}

\textbf{Setup}:
\begin{itemize}
  \item $B = 500$
  \item $\responsedist(\mean) = \text{Bernoulli}(\mean), \link(\mean) = \log(\mean/(1-\mean))$.
  \item $m = 1000, 2000, 5000, 10000$.
  \item $n_1,\ldots,n_m$: computed using \texttt{sample(2:(2 * (n - 1)), size = m, replace = TRUE)} with $n = 3$.
  \item $\sigma_u = 1$.
  \item $k = 5, 9 ,15$.
  \item $p = 1, f(x) = \sin(2\pi x)$.
\end{itemize}

\begin{figure}[p]
  \centering
  \includegraphics[width = \plotwidth, height = \plotheight]{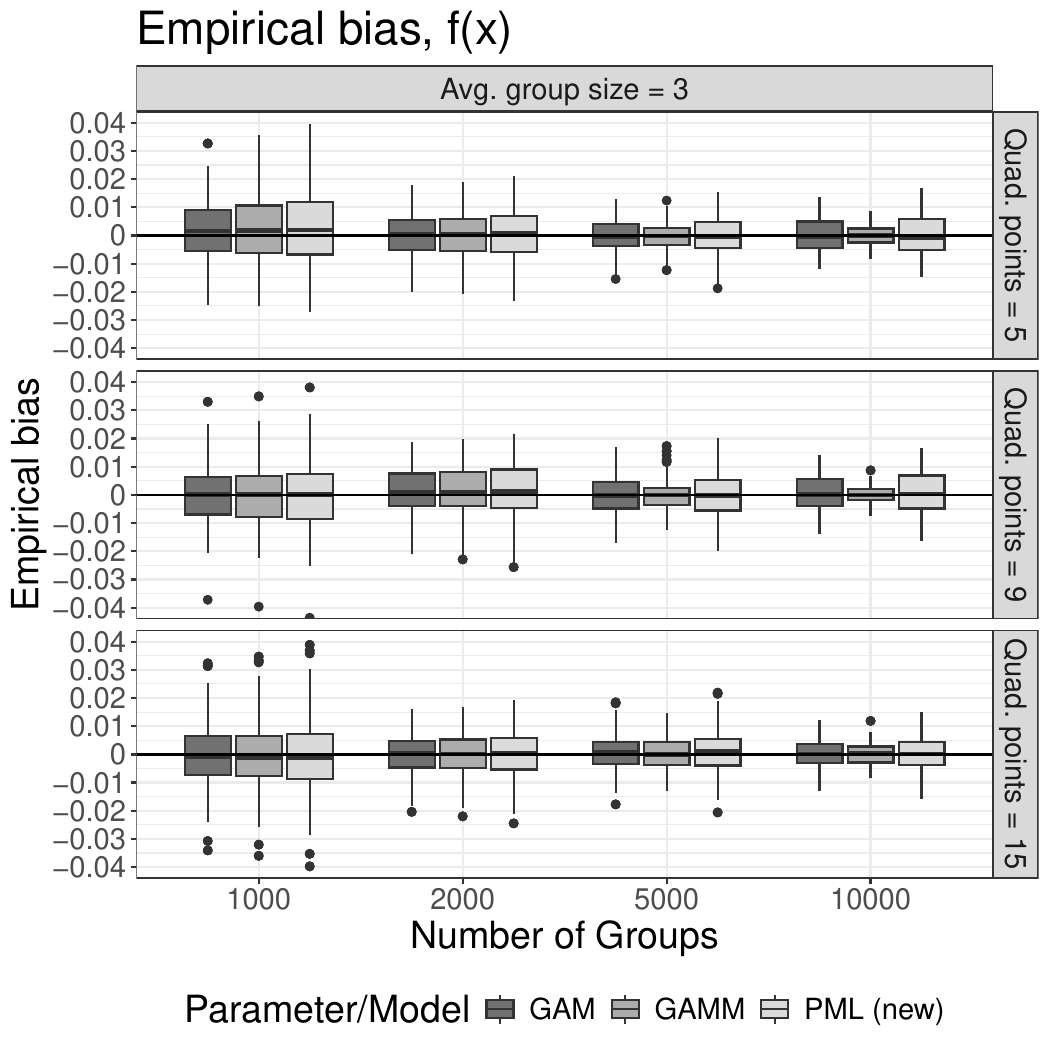}
  \caption{Empirical bias of $\widehat{f}$, true $f(x) = \sin(2\pi x)$, varying $m$ and $k$.}
  \label{fig:kbias}
\end{figure}

\begin{figure}[p]
  \centering
  \includegraphics[width = \plotwidth, height = \plotheight]{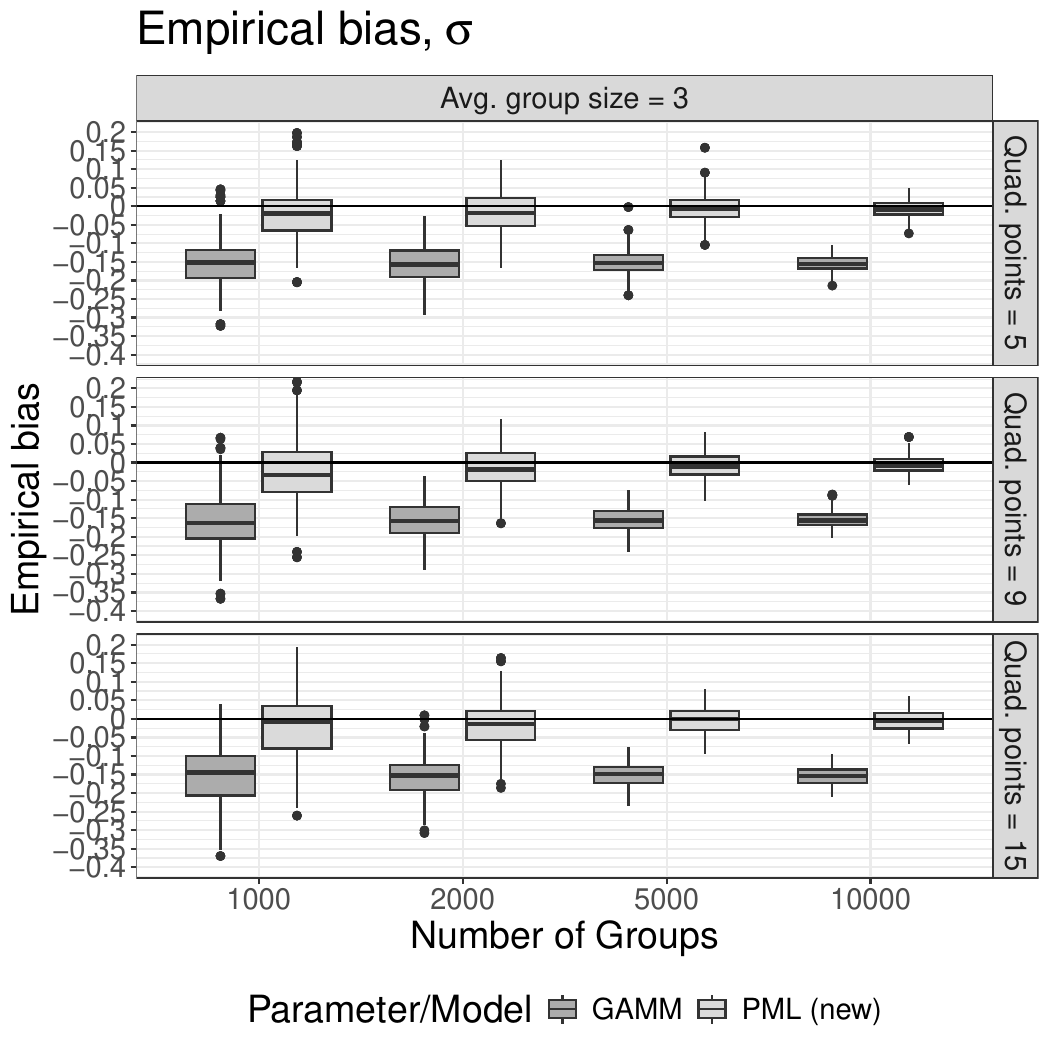}
  \caption{Empirical bias of $\widehat{\sigma}$, true $f(x) = \sin(2\pi x)$, varying $m$ and $k$.}
  \label{fig:ksigma}
\end{figure}

\begin{figure}[p]
  \centering
  \includegraphics[width = \plotwidth, height = \plotheight]{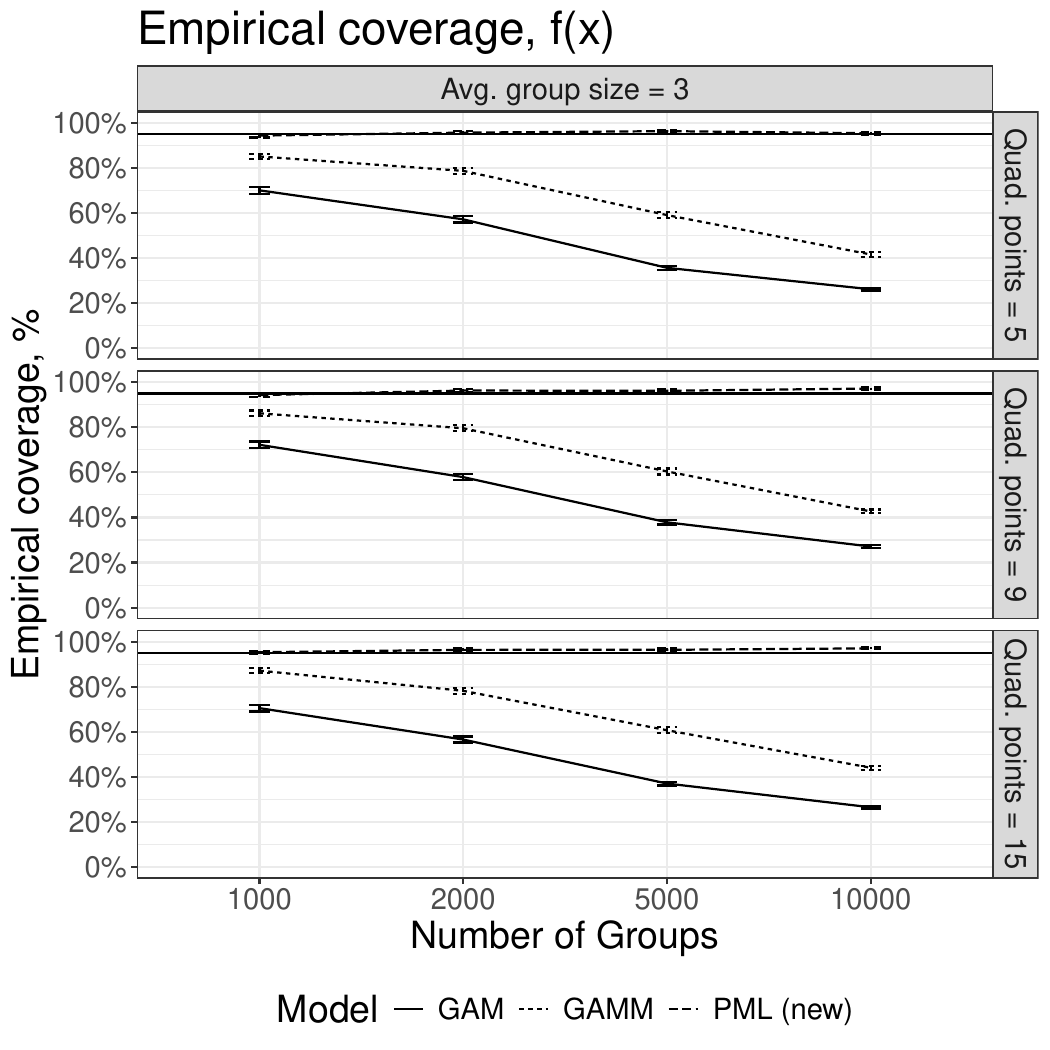}
  \caption{Empirical coverage of $\widehat{f}$, true $f(x) = \sin(2\pi x)$, varying $m$ and $k$.}
  \label{fig:kcovr}
\end{figure}

\textbf{Results}: all three methods have comparable average across-the-function bias of $\widehat{f}$, which is zero on average across the simulations, for each value of $k$ (Figure \ref{fig:kbias}).
The \texttt{gamm} shows non-zero average bias for $\widehat{\sigma}$ which appears to be converging to a nonzero value as $m$ is increased; the new \texttt{pml} method appears to have average bias for $\widehat{\sigma}$ converging to $0$ as $m$ is increased for all values of $k$ (Figure \ref{fig:ksigma}).
The coverage of $\widehat{f}$ for the \texttt{gamm} decreases to far below the nominal level as $m$ is increased, while for \texttt{pml} it appears to level off at a slightly conservative value as $m$ is increased for all values of $k$ (Figure \ref{fig:kcovr}).

\subsection{Very wiggly function}

\textbf{Setup}:
\begin{itemize}
  \item $B = 500$
  \item $\responsedist(\mean) = \text{Bernoulli}(\mean), \link(\mean) = \log(\mean/(1-\mean))$.
  \item $m = 1000, 2000, 5000, 10000$.
  \item $n_1,\ldots,n_m$: computed using \texttt{sample(2:(2 * (n - 1)), size = m, replace = TRUE)} with $n = 3$.
  \item $\sigma_u = 1$.
  \item $k = 5$.
  \item $p = 1, f(x) = (1 / 10)(6g(x; 30, 17) + 4g(x; 3, 11)) - 1$ where $$g(x;\alpha,\beta) = \frac{\Gamma(\alpha + \beta)}{\Gamma(\alpha)\Gamma(\beta)}x^{\alpha-1}(1-x)^{\beta-1}, 0\leq x\leq1, \alpha > 0, \beta > 0$$ is a $\text{Beta}(\alpha,\beta)$ density as defined by \texttt{dbeta} in \texttt{R}; inspired by $f_4$ in the simulation study of \citet{marra2012coverage}.
\end{itemize}

\textbf{Results}: the \texttt{gam} and \texttt{gamm} have comparable average across-the-function bias of $\widehat{f}$, which is zero on average across the simulations but appears
more variable than the case of a simpler true function.
The \texttt{gamm} shows non-zero average bias for $\widehat{\sigma}$ which appears to be converging to a nonzero value as $m$ is increased, an effect which is less severe for the larger $n=9$ than the smaller $n=3$; the new \texttt{pml} method appears to have average bias for $\widehat{\sigma}$ converging to $0$ as $m$ is increased for the larger $n=9$ and nearly to zero for the smaller $n=3$ (Figure \ref{fig:complexsigma}).
The coverage of $\widehat{f}$ for the \texttt{gamm} decreases to far below the nominal level as $m$ is increased, while for \texttt{pml} it appears to level off at a slightly optimistic value as $m$ is increased, an effect which is less severe for the larger $n=9$ than for the smaller $n=3$ (Figure \ref{fig:complexcovr}).

\begin{figure}[p]
  \centering
  \includegraphics[width = \plotwidth, height = \plotheight]{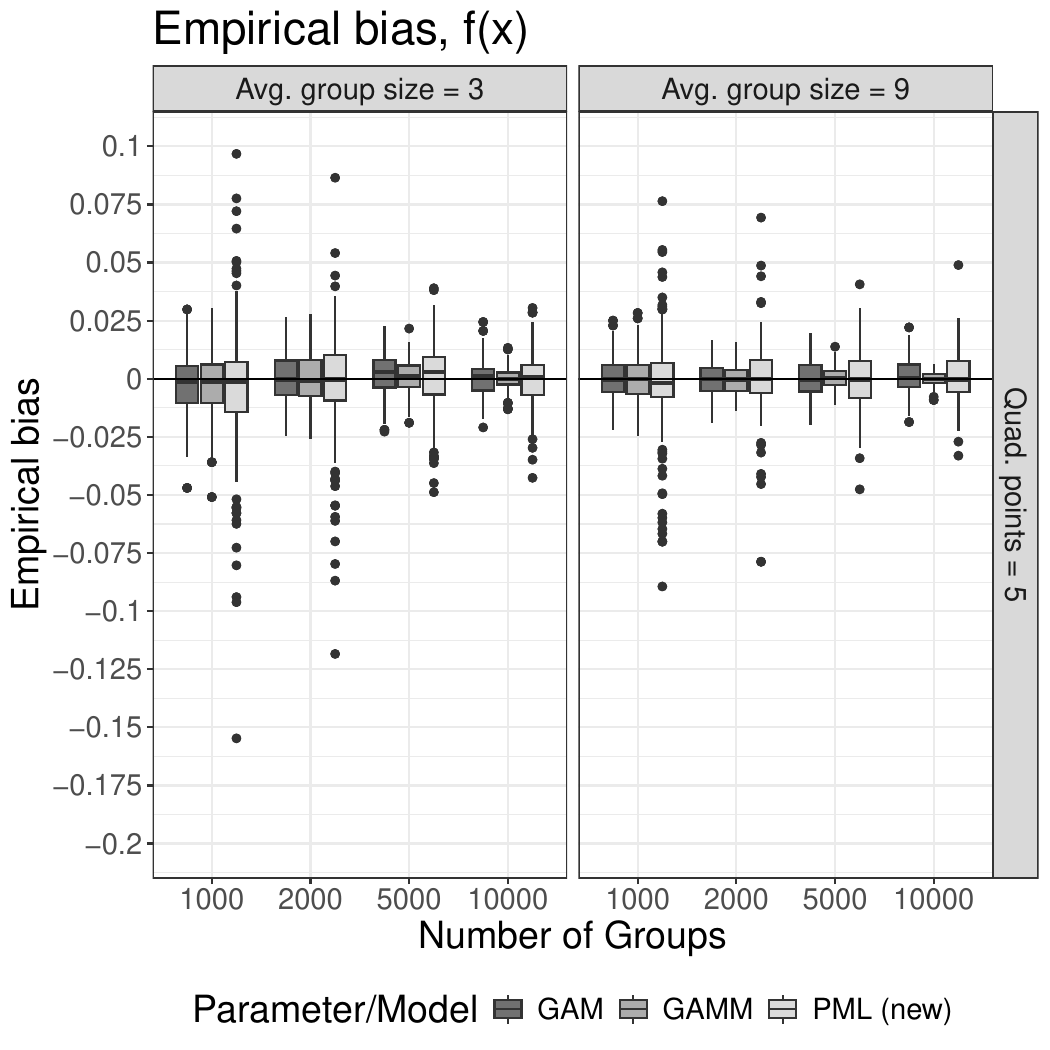}
  \caption{Empirical bias of $\widehat{f}$, true $f(x) = (1 / 10)(6g(x; 30, 17) + 4g(x; 3, 11)) - 1$, varying $m$.}
  \label{fig:complexbias}
\end{figure}

\begin{figure}[p]
  \centering
  \includegraphics[width = \plotwidth, height = \plotheight]{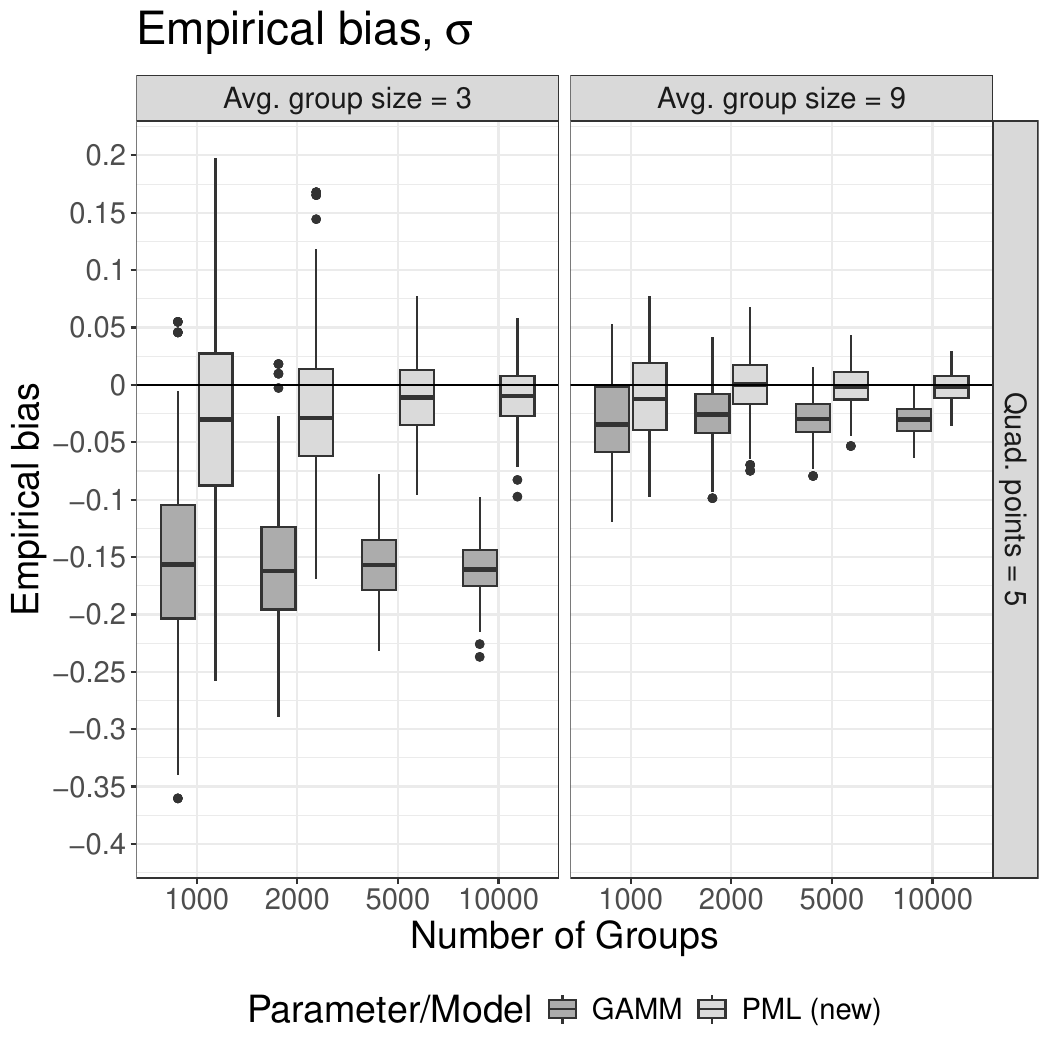}
  \caption{Empirical bias of $\widehat{\sigma}$, true $f(x) = (1 / 10)(6g(x; 30, 17) + 4g(x; 3, 11)) - 1$, varying $m$.}
  \label{fig:complexsigma}
\end{figure}

\begin{figure}[p]
  \centering
  \includegraphics[width = \plotwidth, height = \plotheight]{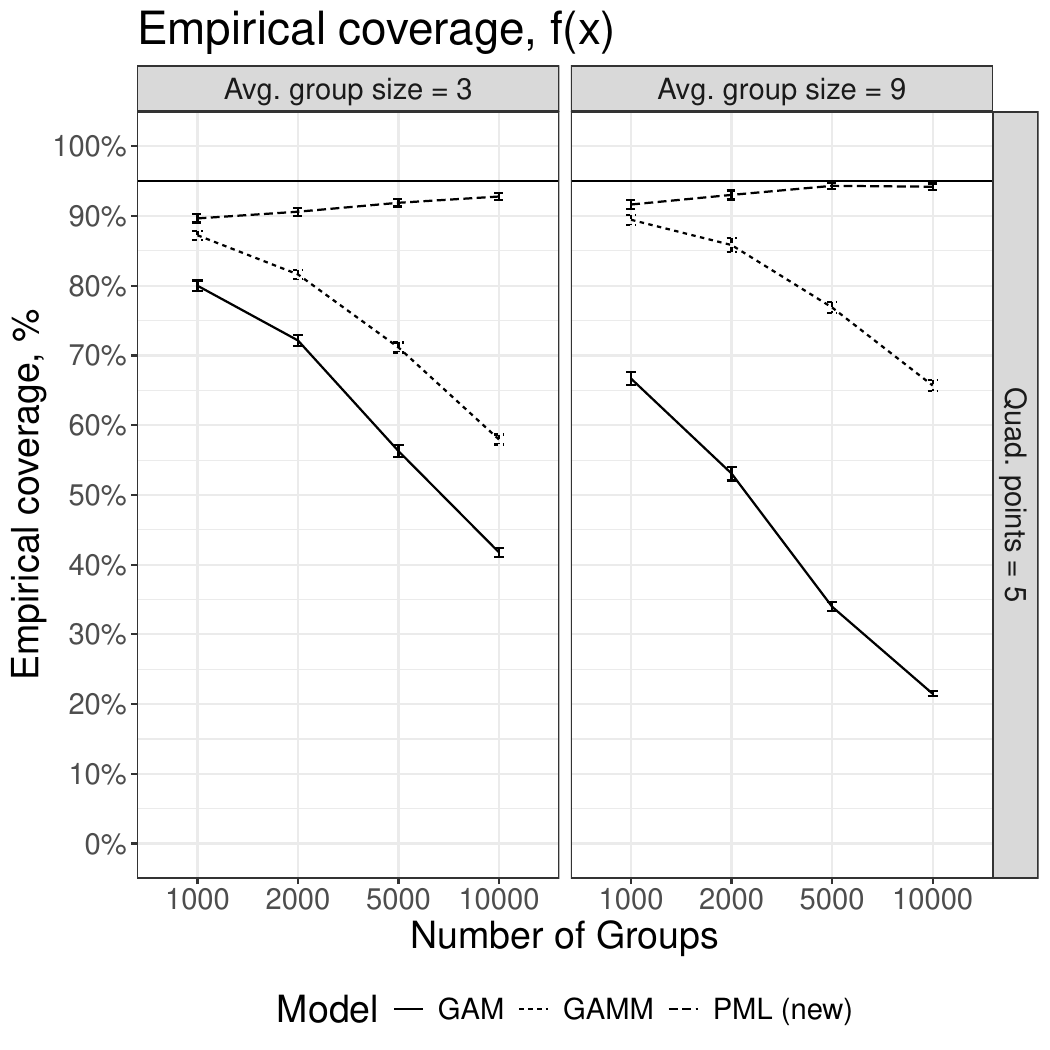}
  \caption{Empirical coverage of $\widehat{f}$, true $f(x) = (1 / 10)(6g(x; 30, 17) + 4g(x; 3, 11)) - 1$, varying $m$.}
  \label{fig:complexcovr}
\end{figure}

\subsection{Poisson response with \texttt{gamm4}}

\textbf{Setup}:
\begin{itemize}
  \item $B = 500$
  \item $\responsedist(\mean) = \text{Poisson}(\mean), \link(\mean) = \log\mean$.
  \item $m = 1000, 2000, 5000, 10000$.
  \item $n_1,\ldots,n_m$: computed using \texttt{sample(2:(2 * (n - 1)), size = m, replace = TRUE)} with $n = 3$.
  \item $\sigma_u = 1$.
  \item $p = 1, f(x) = \sin(2\pi x) + \alpha$ for $\alpha = -2, 0$.
\end{itemize}

A Poisson generalized additive mixed model was fit with linked mean equal to $\link\{\mean(x, u)\} = f(x) + \alpha + u$ for $\alpha = -2, 0, 2$.
Higher $\alpha$ gives a higher mean which is conjectured to lead to a more
accurate Laplace approximation and hence less error when fitting the model using \texttt{gamm4}.
A \texttt{pml} implementation is not available for the Poisson distribution, so this simulation serves only to investigate 
whether the problem that has been observed empirically to occur
with the Bernoulli distribution also seems to occur with the Poisson, not whether the proposed \texttt{pml} approach mitigates the problem.

\begin{figure}[p]
  \centering
  \includegraphics[width = \plotwidth, height = \plotheight]{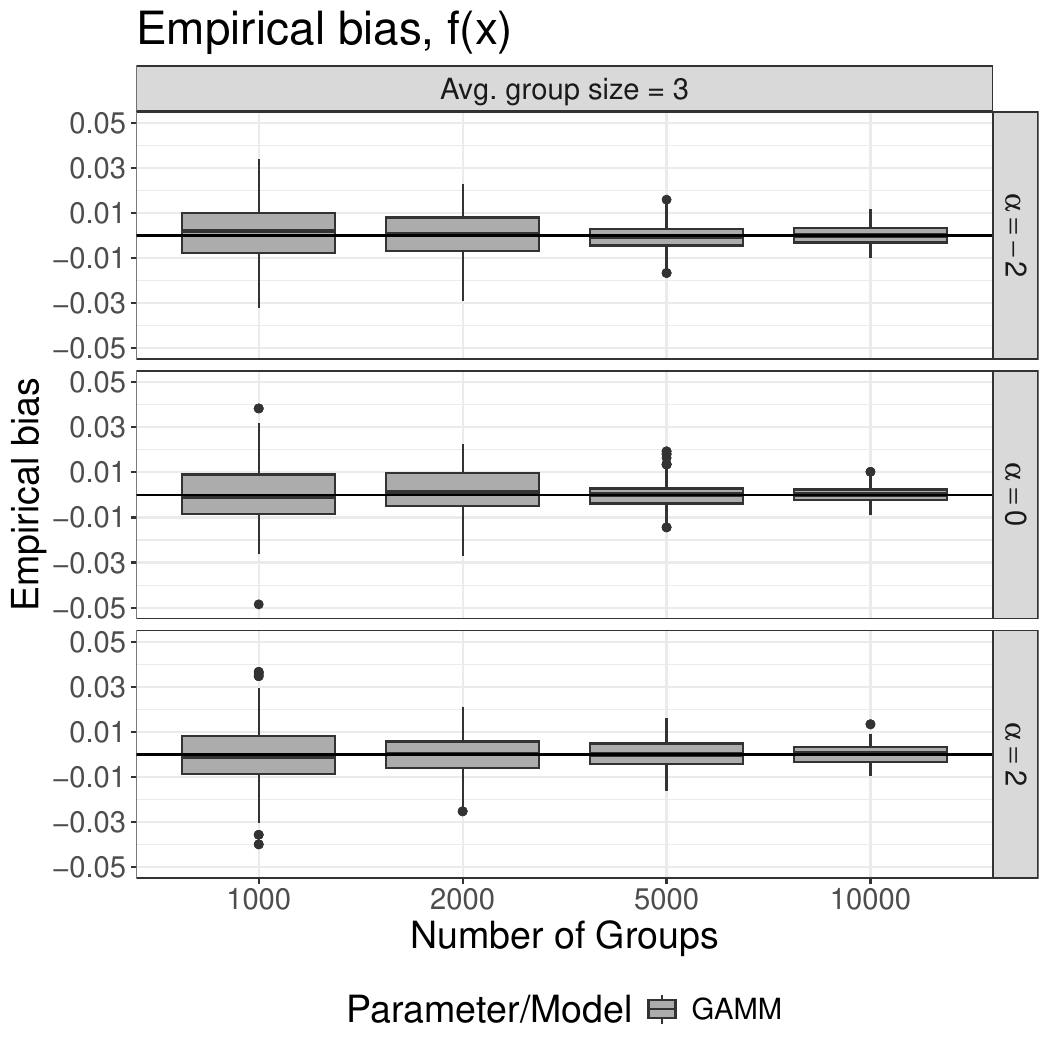}
  \caption{Empirical bias of $\widehat{f}$, true $f(x) = \sin(2\pi x) + \alpha$ for $\alpha = -2, 0$, varying $m$.}
  \label{fig:poissonbias}
\end{figure}

\begin{figure}[p]
  \centering
  \includegraphics[width = \plotwidth, height = \plotheight]{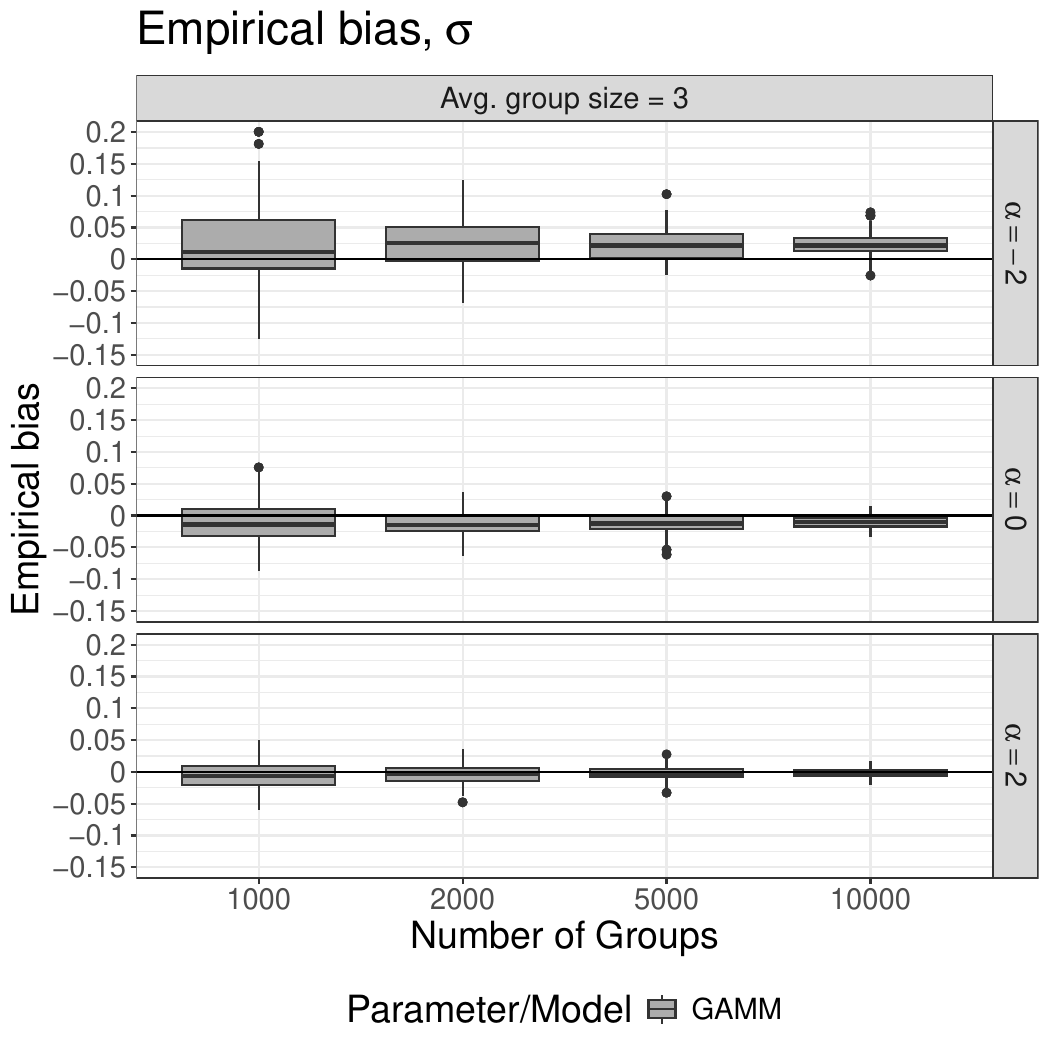}
  \caption{Empirical bias of $\widehat{\sigma}$, true $f(x) = \sin(2\pi x) + \alpha$ for $\alpha = -2, 0$, varying $m$.}
  \label{fig:poissonsigma}
\end{figure}

\begin{figure}[p]
  \centering
  \includegraphics[width = \plotwidth, height = \plotheight]{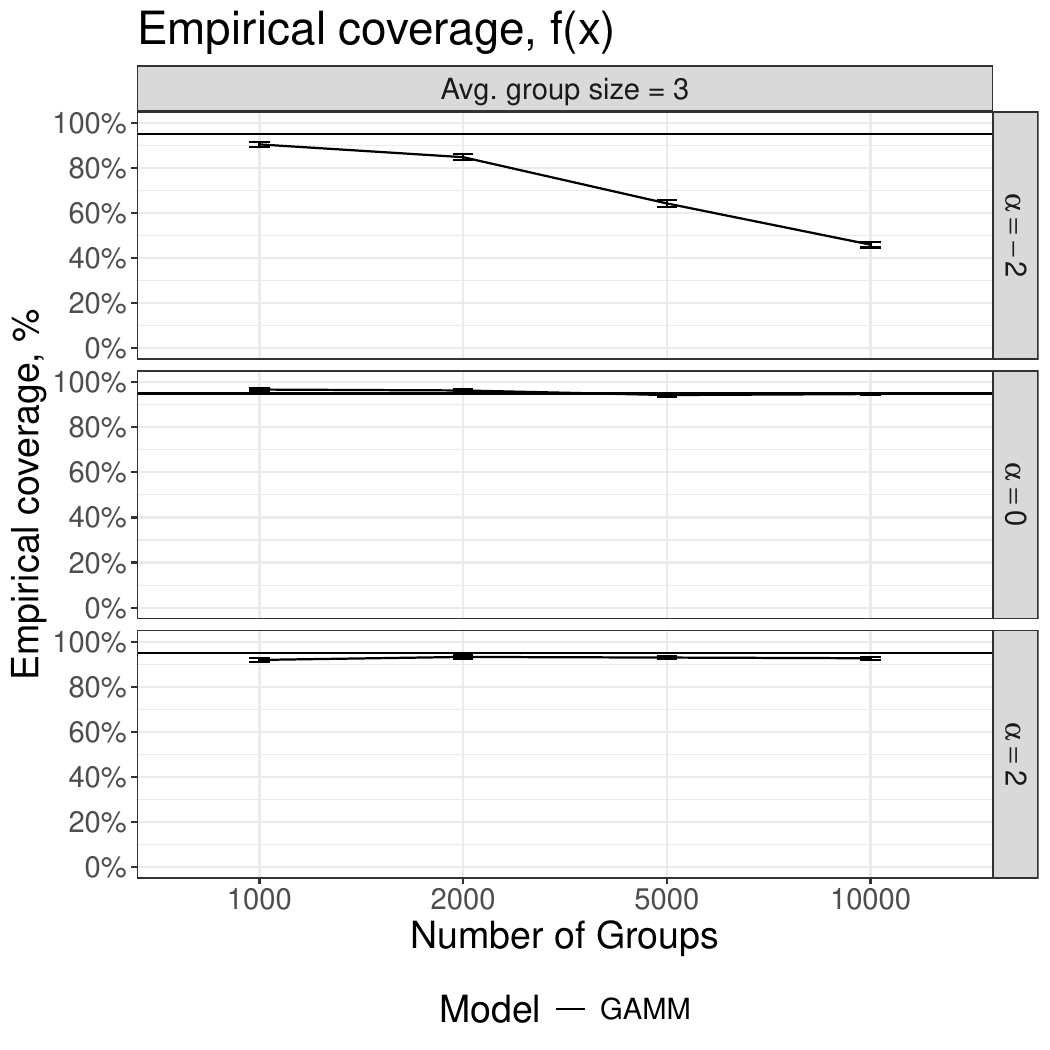}
  \caption{Empirical coverage of $\widehat{f}$, true $f(x) = \sin(2\pi x) + \alpha$ for $\alpha = -2, 0$, varying $m$.}
  \label{fig:poissoncovr}
\end{figure}

\textbf{Results}: for all values of $\alpha$ tried, the \texttt{gamm} has zero average bias (Figure \ref{fig:poissonbias}).
The bias of $\widehat{\sigma}$ appears to converge to $0$
for the highest $\alpha = 2$, and to a value different than zero for $\alpha = 0,-2$ with the problem being most severe at the smallest $\alpha = -2$ (Figure \ref{fig:poissonsigma}).
The coverage appears close to nominal for the larger $\alpha = 0, 2$,
but the problem of low coverage observed with the Bernoulli 
distribution in all the other simulations appears to occur with the lowest ($\alpha = -2$) mean Poisson distribution tried (Figure \ref{fig:poissoncovr}).

\subsection{Small $m$}

\textbf{Setup}:
\begin{itemize}
  \item $B = 500$.
  \item $\responsedist(\mean) = \text{Bernoulli}(\mean), \link(\mean) = \log(\mean/(1-\mean))$.
  \item $m = 100, 200, 500$.
  \item $n_1,\ldots,n_m$: computed using \texttt{sample(2:(2 * (n - 1)), size = m, replace = TRUE)} with $n = 3, 9$.
  \item $\sigma_u = 1$.
  \item $k = 5$.
  \item $p = 1, f(x) = \sin(2\pi x)$.
\end{itemize}

\begin{figure}[p]
  \centering
  \includegraphics[width = \plotwidth, height = \plotheight]{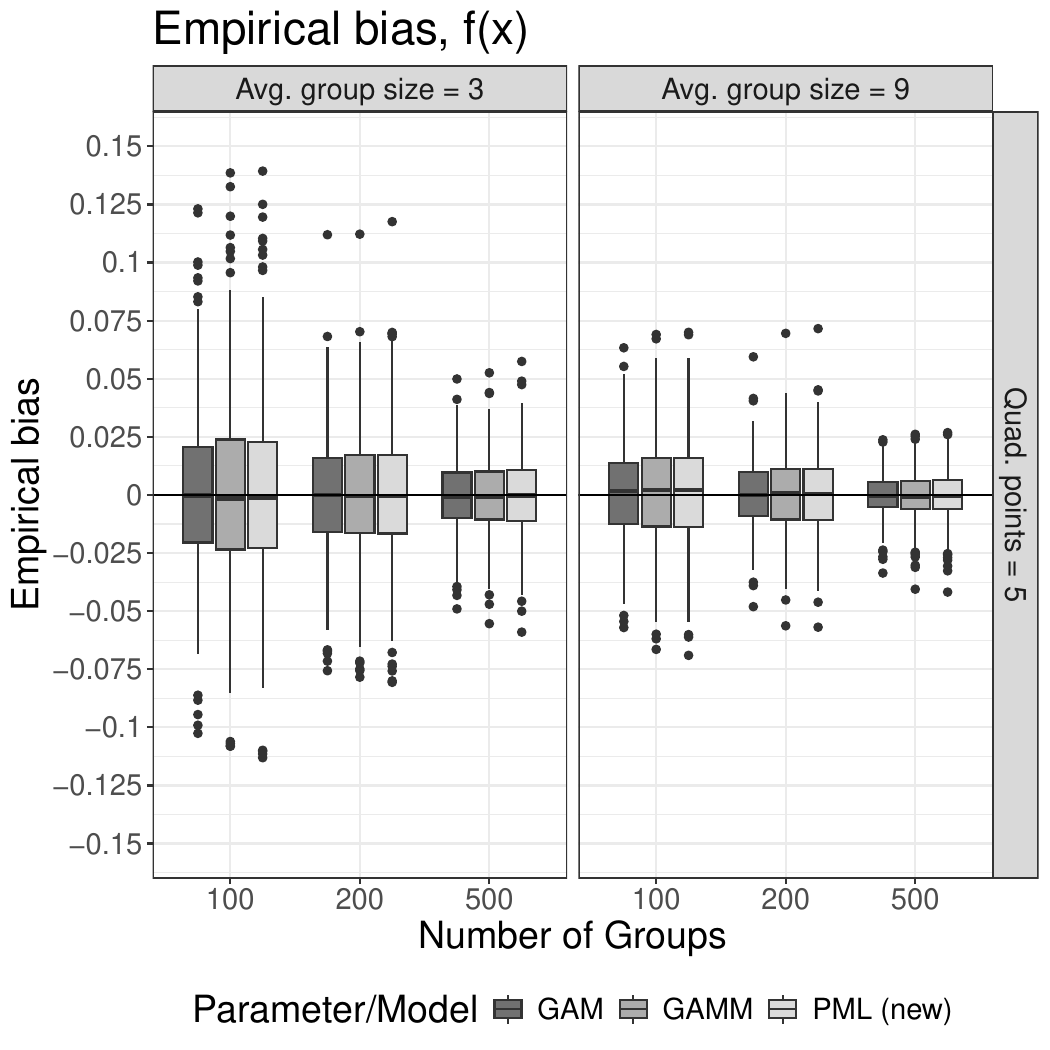}
  \caption{Empirical bias of $\widehat{f}$, true $f(x) = \sin(2\pi x)$, varying $m$ and $n$.}
  \label{fig:smallbias}
\end{figure}

\begin{figure}[p]
  \centering
  \includegraphics[width = \plotwidth, height = \plotheight]{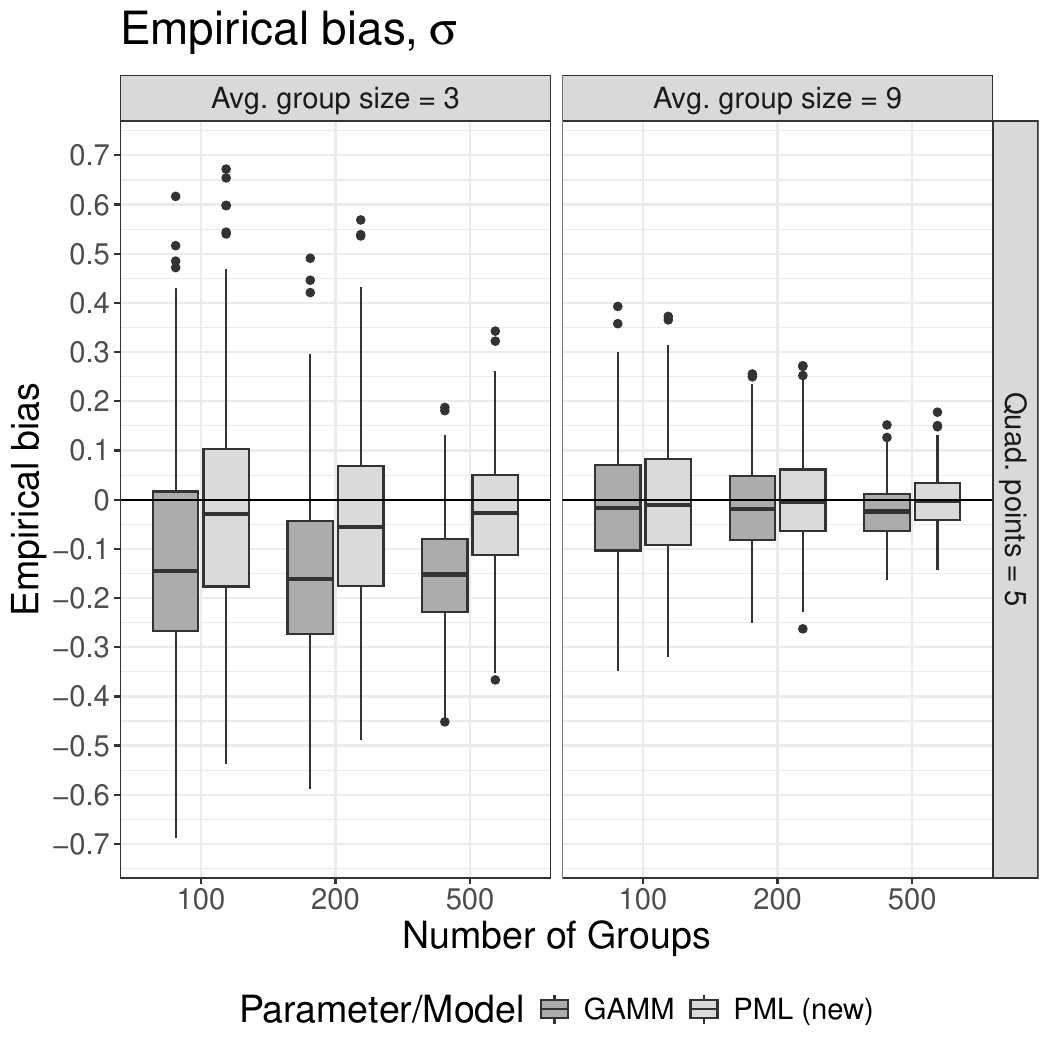}
  \caption{Empirical bias of $\widehat{\sigma}$, true $f(x) = \sin(2\pi x)$, varying $m$ and $n$.}
  \label{fig:smallsigma}
\end{figure}

\textbf{Results}: all three methods have comparable average across-the-function bias of $\widehat{f}$, which is zero on average across the simulations (Figure \ref{fig:smallbias}).
The \texttt{gamm} and \texttt{pml} estimates both show non-zero average bias for $\widehat{\sigma}$ at $n=3$ which appears to be converging to a nonzero value as $m$ is increased,
an effect which appears less severe for \texttt{pml}
than \texttt{gamm}. For $n=9$, both \texttt{gamm} and \texttt{pml} yield nearly zero average bias for $\sigma$
(Figure \ref{fig:smallsigma}).
The coverage of $\widehat{f}$ for the \texttt{gamm} is lower than nominal for $n=3$ and nominal for $n=9$ and does not change predictably for increasing $m$. The coverage of
$\widehat{f}$ for \texttt{pml} does increase with increasing
$m$ and reaches nominal for $n=9$ and just below nominal for $n=3$.

\begin{figure}[p]
  \centering
  \includegraphics[width = \plotwidth, height = \plotheight]{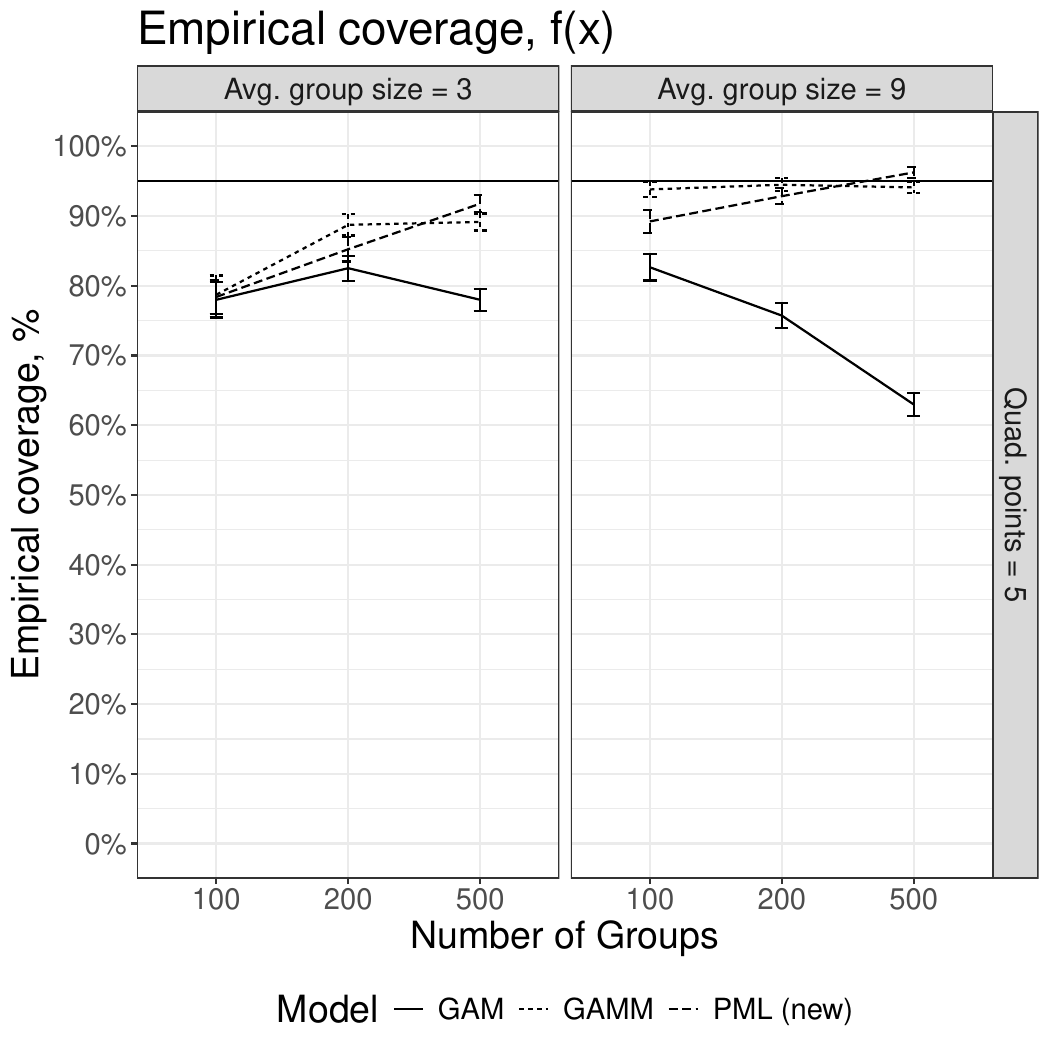}
  \caption{Empirical coverage of $\widehat{f}$, true $f(x) = \sin(2\pi x)$, varying $m$ and $n$.}
  \label{fig:smallcovr}
\end{figure}

\subsection{Flat $f$}

\textbf{Setup}:
\begin{itemize}
  \item $B = 1000$.
  \item $\responsedist(\mean) = \text{Bernoulli}(\mean), \link(\mean) = \log(\mean/(1-\mean))$.
  \item $m = 100, 200, 500$.
  \item $n_1,\ldots,n_m$: computed using \texttt{sample(2:(2 * (n - 1)), size = m, replace = TRUE)} with $n = 3, 9$.
  \item $\sigma_u = 1$.
  \item $k = 25$.
  \item $p = 1, f(x) = 2x - 1$.
\end{itemize}

\begin{figure}[p]
  \centering
  \includegraphics[width = \plotwidth, height = \plotheight]{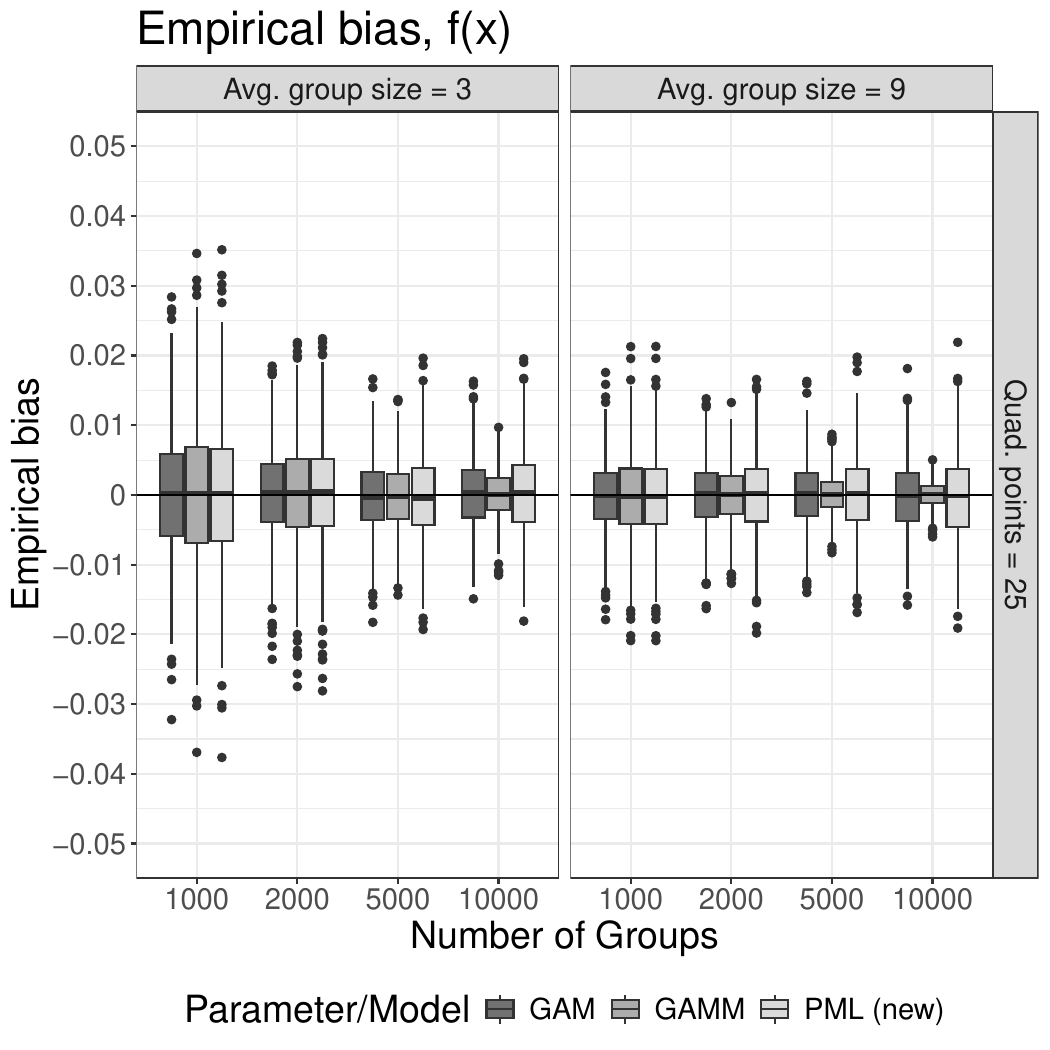}
  \caption{Empirical bias of $\widehat{f}$, true $f(x) = 2x-1$, varying $m$ and $n$.}
  \label{fig:flatbias}
\end{figure}

\begin{figure}[p]
  \centering
  \includegraphics[width = \plotwidth, height = \plotheight]{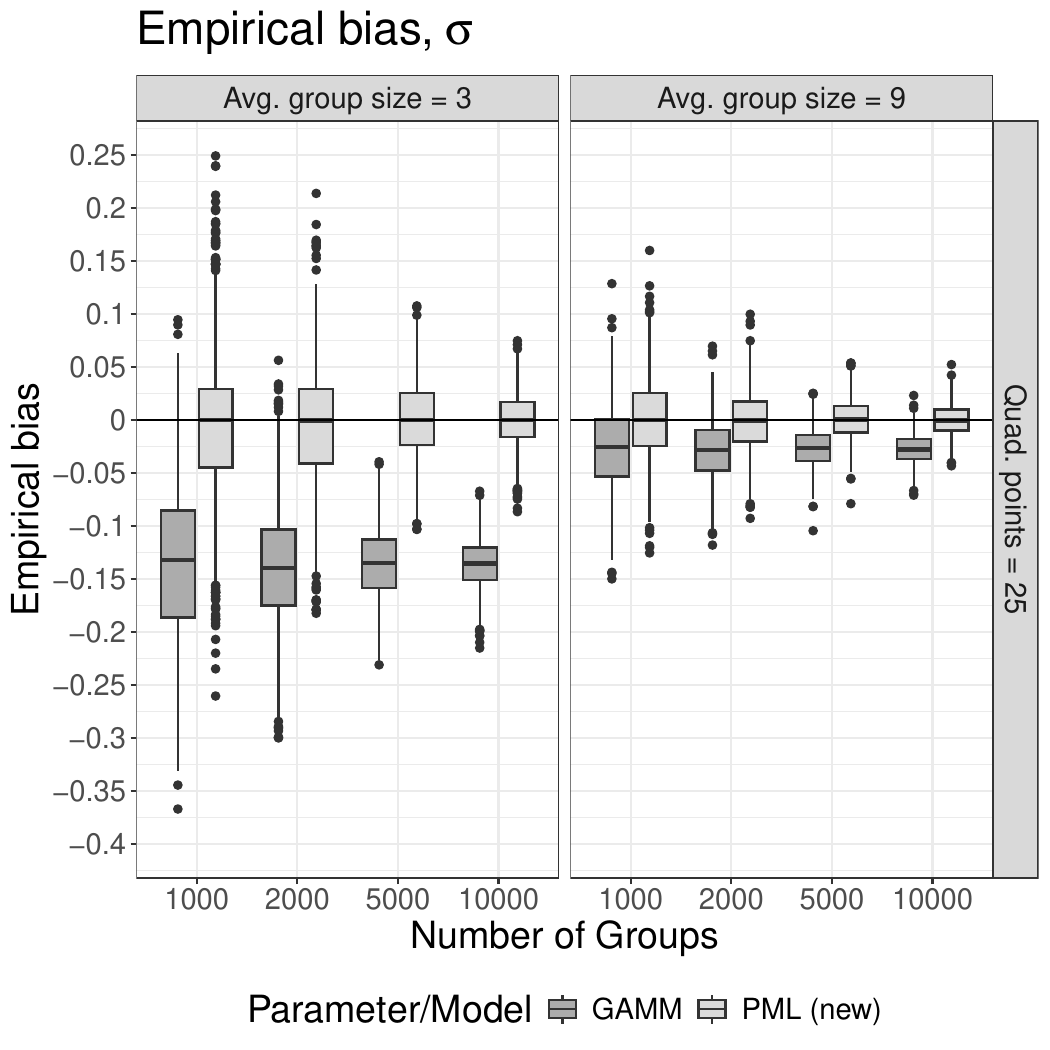}
  \caption{Empirical bias of $\widehat{\sigma}$, true $f(x) = 2x-1$, varying $m$ and $n$.}
  \label{fig:flatsigma}
\end{figure}

\begin{figure}[p]
  \centering
  \includegraphics[width = \plotwidth, height = \plotheight]{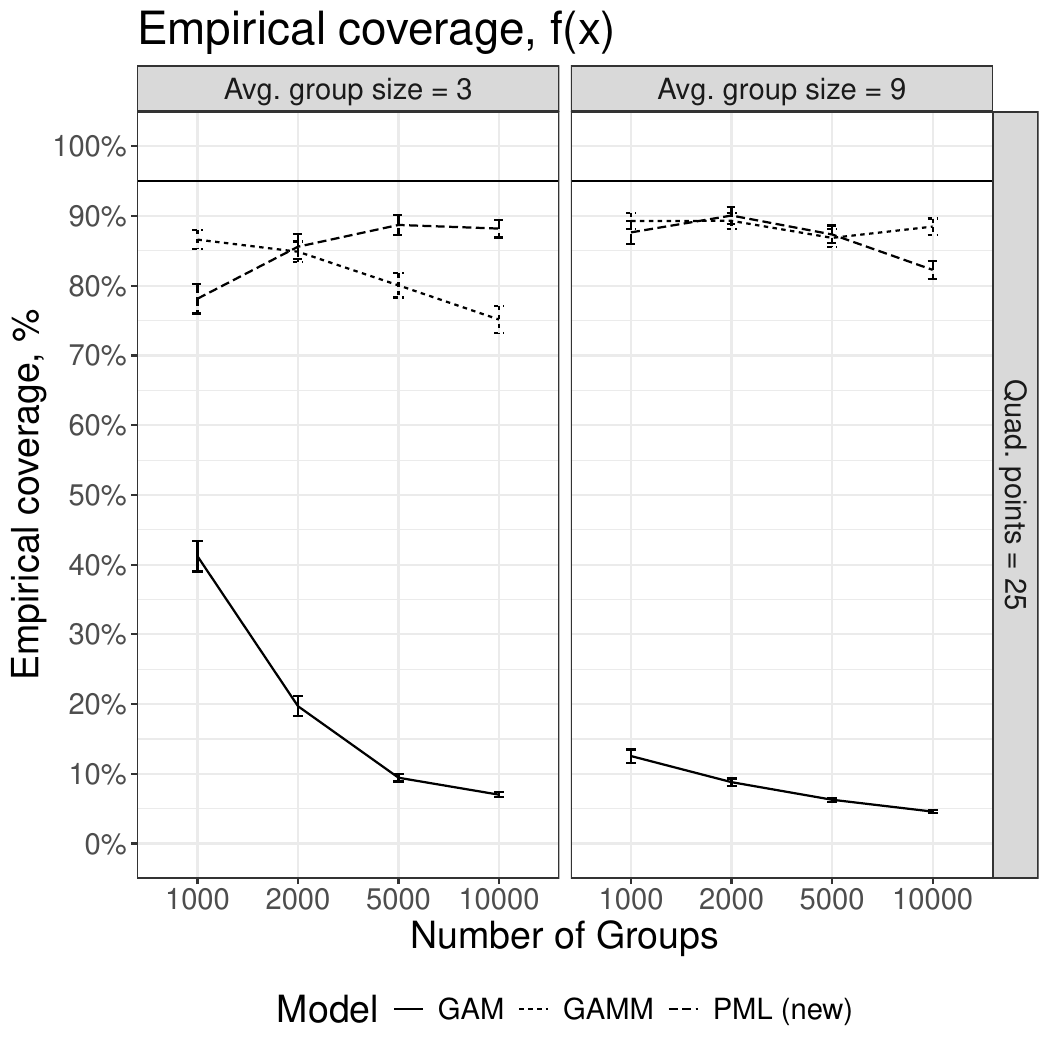}
  \caption{Empirical coverage of $\widehat{f}$, true $f(x) = 2x-1$, varying $m$ and $n$.}
  \label{fig:flatcovr}
\end{figure}

\textbf{Results}: all three methods have comparable average across-the-function bias of $\widehat{f}$, which is zero on average across the simulations (Figure \ref{fig:flatbias}).
The \texttt{gamm}shows non-zero average bias for $\widehat{\sigma}$ which appears to be converging to a nonzero value as $m$ is increased.
This value is closer to zero for $n=9$ than for $n=3$.
The $\texttt{pml}$ method
attains zero average bias for $\sigma$ for all values of $m$ and $n$.
(Figure \ref{fig:flatsigma}).
The coverage of $\widehat{f}$ for the \texttt{gamm} is lower than nominal for $n=3$ and decreases with increasing $m$. 
The coverage of
$\widehat{f}$ for \texttt{pml} is also too low, in some cases comparable to the \texttt{gamm}, and in some cases better and some worse (Figure \ref{fig:flatcovr}); this behaviour occurs even with a high number $k=25$ of quadrature points.
When $f$ is linear, the ``true'' $\lambda=\infty$ and it is expected that its estimation will be challenging.
This simulation shows that the \texttt{pml} method does not address the low coverage problem in a case when estimation of $\lambda$
is expected to be challenging.

\end{document}